\documentclass[twocolumn]{aastex702}
\usepackage{multirow}
\usepackage{xspace}
\usepackage{gensymb}
\usepackage{subcaption}

\usepackage{graphicx}
\usepackage{amsmath}

\renewcommand{\arraystretch}{2}

\newcommand{\buv}{\,{\beta_{\rm UV}}}
\newcommand{\buvmod}{\,{\beta_{\rm UV}^{\rm mod}}}
\newcommand{\buvobs}{\,{\beta_{\rm UV}^{\rm obs}}}
\newcommand{\bopt}{\,{\beta_{\rm opt}}}

\newcommand{\Mstar}{M_\star}
\newcommand{\Ha}{\rm H\alpha}
\newcommand{\Hb}{\rm H\beta}

\begin{document}

\title{Host Galaxy UV Emission and Dust Reddening in Little Red Dots}

\correspondingauthor{Emmanuel Durodola}
\email{emmanuel.a.durodola.gr@dartmouth.edu}

\author[0009-0004-9516-9593]{Emmanuel Durodola}
\affiliation{Department of Physics \& Astronomy, Dartmouth College, 6127 Wilder Laboratory, Hanover, NH 03755, USA}
\email[]{emmanuel.a.durodola.gr@dartmouth.edu}

\author[0000-0003-1468-9526]{Ryan C. Hickox}
\affiliation{Department of Physics \& Astronomy, Dartmouth College, 6127 Wilder Laboratory, Hanover, NH 03755, USA}
\email[]{ryan.c.hickox@dartmouth.edu}

\author[0000-0003-1420-6037]{Jonathan Cohn}
\affiliation{ Department of Physics
Honors College 701 W Grace Street
Box 842041
Richmond, VA 23284, USA}
\email[]{cohnj3@vcu.edu}

\author[0000-0002-5896-6313]{David M. Alexander}
\affiliation{Centre for Extragalactic Astronomy, Department of Physics. Durham University, South Road, Durham, DH1 3LE, UK}
\email[]{d.m.alexander@durham.ac.uk}

\begin{abstract}

Little Red Dots (LRDs) have been the subject of extensive analysis to place them within the context of galaxy and black hole evolution. Despite several proposed scenarios for the nature of the rest optical emission in LRDs, our understanding of the UV emission  remains a matter of debate. Here, we aim to study the origin of the UV emission in LRDs. We conduct emission-line and continuum fitting analysis of thirteen LRDs using a combination of NIRSpec PRISM and medium-resolution G235M and G395M spectroscopy. We isolate and analyze the narrow-line emission arising from larger physical scales in LRDs. We observe that the narrow-line Balmer decrements are typically greater than those predicted for case B recombination and broadly correlated with both the UV slope and the break strength, providing clear evidence that the UV emission originates from galaxy scales and its slope is significantly affected by dust reddening. SED fitting analysis further shows that the UV continuum of LRDs is consistent with a dust attenuated stellar template. Finally, we show through a comparison of the UV slope and the Balmer break strength, that LRDs can be further divided based on break strength, such that the canonical LRD spectral shape (strong break strength ($>$ 3) and flat rest UV continuum) represents the most extreme sub-population. Our results imply that for most LRDs, the UV emission originates primarily from the young host galaxy, and that the characteristic optical emission may be due to a combination of dust reddening and intrinsically red accretion processes.

\end{abstract}


\section{Introduction} 
\label{sec:intro}

JWST has revealed a ubiquitous population of compact, red objects at high redshift (z$>$4) that present a significant challenge to our current understanding of galaxy evolution in the early universe. Commonly referred to as Little Red Dots (LRDs: \citet{Matthee_2023}), these objects are selected on the basis of their compact morphologies, peculiar SED shapes, and excess red rest-frame optical emission \citep[e.g.,][]{Kocevski_2023, Maiolino_2023}. Spectroscopic analysis of LRDs has revealed broad Balmer emission lines (FWHM $\geq$ 1000 km $s^{-1}$) \citep{greene_uncover_2023} in $\sim$ 80\% of the population \citep{Hviding+25}, suggesting that LRDs represent a population of high-redshift Type I accreting AGN. SED fitting and emission line analyses have further suggested that LRDs host black hole masses of $\sim$ 10\% of the host galaxy stellar mass, which would make LRDs among the most extreme cases of overmassive black holes currently known \citep[e.g.,][]{pacucci_jwst_2023, Maiolino_2023_new, barro_extremely_2023}. Yet LRDs are notably X-ray weak, measuring 10 - 100 times weaker than local Type I AGN \citep[e.g.,][]{Ananna_2024, Yue_24}, which is in tension with the expected signatures of such massive accreting black holes. LRDs also exhibit red rest-frame optical emission typical of dust-reddened AGN; however, non-detections in the far-infrared and a turnover in the rest-frame near-infrared continuum suggest that this red emission is unlikely to arise from dust attenuation alone \citep{Setton_24_dust}. Furthermore, the space density of LRDs drops off at $z<4$ \citep[e.g.,][]{Ma_26, Kapoor2026}), despite being numerically abundant (number densities $>$ 10$^{-5}$ Mpc$^{-3}$) at z$\geq$4 \citep[e.g.,][]{barro_extremely_2023, Matthee_2023}, suggesting that the physical conditions responsible for the formation of LRDs are more readily achieved in the high-redshift universe.   

Alternatively, the broad emission lines observed in LRDs have been interpreted as arising from the rotational dynamics of massive star-forming galaxies \citep[e.g.,][]{Baggen_2023, Kokubo_25}), particularly given that LRDs exhibit Balmer breaks \citep{Setton_2024} that are typically associated with bound-free absorption in the atmospheres of evolved stars. However, the compact morphologies of LRDs, their high redshifts, and the stellar masses that would be required ($\sim$10$^{10}$ M$_{\odot}$) imply a rate of stellar mass assembly that is difficult to reconcile with the short dynamical timescales available at these epochs \citep[e.g.,][]{Labbe_2023, Baggen_2023, Labbe_24, Wang_24}; it has been proposed that dark matter accelerated star formation in the early universe could result in similar stellar masses and sizes of LRD \citep{Boylan-kolchin_25}. Furthermore, recent spectroscopic observations have revealed Balmer break strengths in some LRDs that exceed the maximum theoretically achievable in star-forming galaxies, suggesting instead that this feature is tied to dense gas enshrouding of a broad-line AGN, a ``black hole star'' in the nucleus of the LRD \citep[e.g.,][]{Naidu_25, de_graff_25a, Taylor_25}, which is postulated to exist in all LRDs \citep{DeGraff_2025}. While the physical origin of the dense gas remains unclear, the presence of a dense gas envelope surrounding the broad-line region (BLR) naturally accounts for many of the spectral properties of LRDs, the X-ray quietness, an intrinsically red rest-frame optical continuum, why LRDs remain undetected in the far-infrared. This picture is further supported by higher resolution spectroscopic observations of LRDs that have revealed absorption features in the Balmer emission lines, interpreted as evidence of gas outflows or inflows in the nuclear regions \citep[e.g.,][]{Ji_2025, Matthee_2023, juodzbalis_24, Deugenio_25}.  

Despite the ability of the ``black hole star'' model to reproduce many of the spectral features observed in LRDs, particularly in the rest-frame optical, it presents a challenge in explaining the faint UV emission. Within this framework, the dense gas enshrouding the AGN is expected to absorb all emission blueward of the Balmer break; however, a faint red UV continuum is unambiguously detected in LRDs \citep{Kokorev2024, Hainline2025, Kocevski_2024}. Some studies have interpreted this as evidence that the covering fraction of the gas envelope is not unity, such that a fraction of intrinsically red scattered AGN emission is able to escape, giving rise to the observed UV emission \citep{Taylor_25, Pacucci_26}. Alternatively, this could be evidence that the emission from LRDs is a composite, with a black hole star in the nucleus of faint star forming host galaxies \citep{Naidu_25, DeGraff_2025}. Morphological analysis of the rest frame UV photometry have suggested that some LRDs have extended components indicative of a host galaxy \citep{Rinaldi_2025, Cloonan_26_extended_uv_LRD}. Furthermore, the steep Balmer breaks that represent the primary observational evidence for the black hole star model have only been observed in a small subset of LRDs, and the associated Balmer absorption feature is estimated to be present in only $\sim$ 40\% of LRDs \citep{Inayoshi_Maiolino_2025}\footnote{} raising the question of whether this model can be applied to the LRD population as a whole.

There is now a growing consensus that LRDs do not constitute a monolithic population, as recent studies have demonstrated that LRDs exhibit diverse spectral shapes \citep{kokorev_25, Barro_25, Perez-gonzalez_26}. Some LRDs display a Balmer break and faint UV emission alongside a flattened rest-frame near-infrared continuum, while others feature a strong blue UV continuum, power-law spectra without a Balmer break, and a continuum rise towards the rest-frame near-infrared. Efforts to explain this diversity have led to composite model interpretations in which the enshrouded black hole, which dominates the rest-frame optical emission, resides in host galaxies of varying stellar mass, giving rise to the observed spectral variation. The rise in the rest-frame near-infrared continuum has been interpreted as evidence of AGN-heated dust \citep{Delvechio_25}, although this could alternatively be attributed to nuclear attenuation of the enshrouded black hole or to intrinsic variations in its luminosity \citep{Barro_25}. Notably, however, it remains unclear whether the host galaxy itself is significantly dust-attenuated.

SED fitting analyses of LRDs using composite templates favor the presence of large-scale dust attenuation \citep{Durodola_25, Rinaldi_2025, Billand_26}. However, Balmer decrement measurements have not provided a clear and consistent picture of this effect. The broad-line Balmer decrement in LRDs deviates strongly from case B recombination, likely due to contributions from the broad-line AGN and scattering effects within the dense gas envelope of the enshrouded black hole model. Under the host galaxy interpretation, the narrow-line Balmer decrement should serve as a reliable tracer of large-scale dust attenuation; and indeed, previous analyses have reported narrow-line Balmer decrement values that, while elevated, are broadly consistent with case B recombination \citep{killi_24_lrd_spectrum, de_graff_25a, Billand_26}. However, these analyses have been limited to individual follow-up observations of single objects, making it difficult to establish a systematic relationship between the narrow-line Balmer decrement and the multiwavelength properties of LRDs. In this analysis, we compile a sample of medium-resolution spectroscopic observations of LRDs with coverage of the Balmer emission lines to measure the narrow-line Balmer decrement and investigate how large-scale dust attenuation in LRDs influences their multiwavelength properties. In Section \ref{sec:data}, we describe our data and sample selection. In Section \ref{sec:method}, we describe our continuum and Balmer line fits. Section \ref{sec:result} displays our results from the fitting, including UV slope, narrow line Balmer decrements, Balmer break strength, and optical continuum slopes. In Section \ref{sec:conclusions}, we discuss the implications of our results on our understanding of LRDs. Finally, in Section \ref{sec:summary} presents a summary of this work.

\section{Data}
\label{sec:data}

This study is based on the sample of LRDs presented in \citet{DeGraff_2025}. The LRDs were selected by simultaneously fitting broken power laws to NIRSpec PRISM spectra, identifying objects with a spectral break near the Balmer limit (H$_\infty$) that satisfy the following criteria: a rest-frame optical slope $\bopt$ $>$ 0, the UV slope $\buv$ $<$ -0.2, and a color difference $\bopt$ - $\buv$ $>$ 0.5, and a compactness criterion requiring either that the ratio of the F444W flux measured within 0\farcs 2 and 0\farcs1 apertures is $<$ 1.7, or that the point source component contributes $>$ 50\% of the total F444W flux. They select 116 unique LRDs. We download the 1D PRISM spectra from the DAWN JWST Archive (DJA; \citealt{brammer_2023_8319596}), and compile complimentary NIRCam photometry by cross-matching the sample with archival JWST photometry from ASTRODEEP \citep{Merlin_24_astrodeep}. The photometric catalog encompasses observations from multiple JWST fields; JADES-GS, JADES-GN, CEERS, ABELL2744, PRIMER-EGS, PRIMER-COSMOS, and NGDEEP. After cross-matching, we recover 89 unique LRDs with both PRISM spectra and NIRCam photometry.

\begin{figure*}
    \centering
    \includegraphics[width=\linewidth]{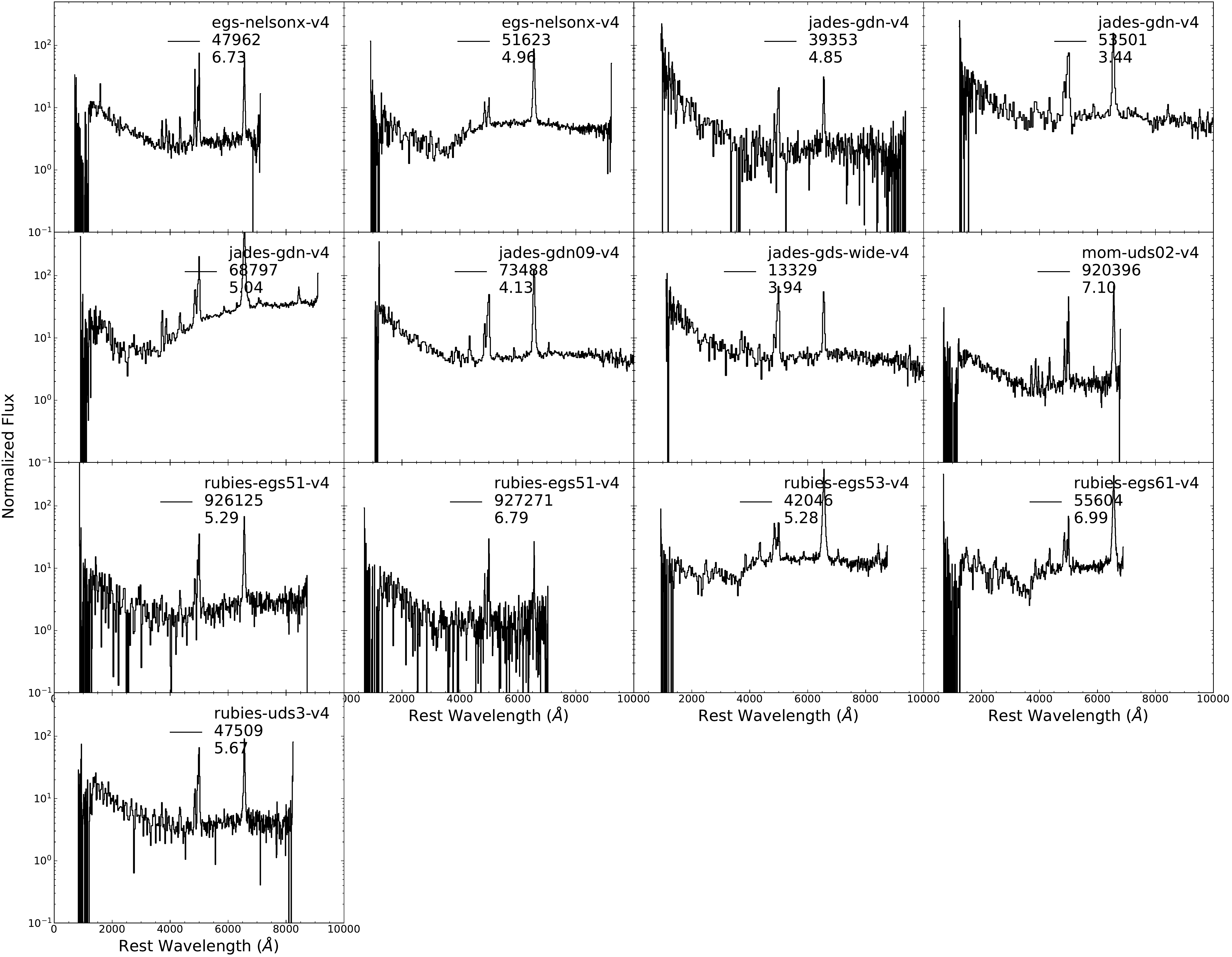}
    \caption{JWST NIRSpec PRISM spectra of the 13 LRDs with medium-resolution counterpart observations. The sub-population contains five LRDs from the RUBIES survey, one from the Mirage-Or-Miracle (MOM) dataset, five from the JADES survey, and two from the EGS-NELSONX dataset.}
    \label{fig:prism_spec}
\end{figure*}

Using DJA, we perform a systematic search for medium-resolution NIRSpec observations of the 89 LRDs in our sample. We identify a subsample of 16 LRDs with medium-resolution (G395M or G235M) grating observations of H$\alpha$ and H$\beta$. We further refine the sample by excluding three LRDs: two lacking spectroscopic coverage in the rest-frame UV, where we measure the observed UV continuum slope, $\buv$, and one exhibiting a Balmer absorption feature that coincides with the line centroid, making reliable constraints on the narrow emission line component impossible. This yields a final medium-resolution subsample of 13 LRDs, which are shown in Figure \ref{fig:prism_spec}.

\begin{figure*}[!ht]
    \centering
    \includegraphics[width=\linewidth]{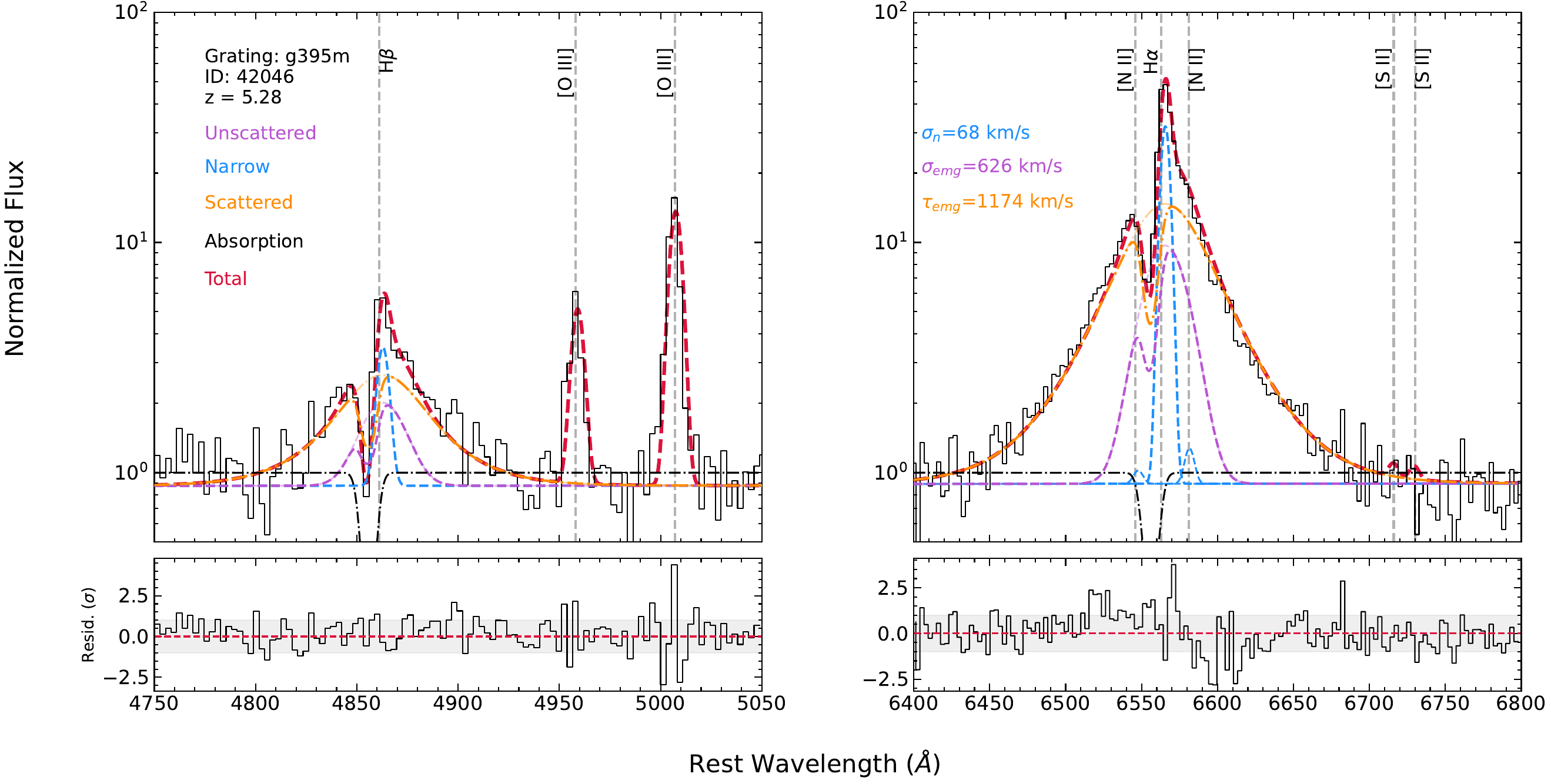}
    \caption{ Example of the two-component Gaussian + exponential emission line fitting of $\Hb$ (left) and $\Ha$ (right) for one of our medium-resolution LRDs: RUBIES-42046. The orange dashed line depicts the scattered emission (an intermediate Gaussian convolved with a symmetric exponential), the purple dashed line depicts the unscattered intermediate Gaussian, the blue dashed line depicts the narrow emission, the black dashed line depicts the narrow Balmer absorption, and the red dashed line depicts the total model fit to the data. Vertical dashed line markers indicate the positions of the Balmer emission lines alongside the forbidden emission lines included in the fitting analysis: [\ion{O}{3}] doublet, [\ion{N}{2}] doublet, and [\ion{S}{2}] doublet. The residuals have also been added in a separate panel. The best-fit values for the different velocity width of each line comp are indicated. }
    \label{fig:cutout}
\end{figure*}

\section{Methodology}
\label{sec:method}

We model the UV continuum as a power-law of the form $f_{\lambda}\propto \lambda^{\buv}$ \citep{Calzetti_94_uv_slope}, where $\buv$ denotes the power law slope of the UV continuum, allowing us to quantify the spectral slope of the UV continuum and to compare it with optical line ratios to study dust attenuation effects on the UV emission.

To measure the optical line ratios, we fit the Balmer emission using a two component Gaussian profile convolved with an exponential profile implemented within the Bayesian modeling framework \texttt{PYSTAN} \citep{pystan}. We simultaneously fit the H$\alpha$ and H$\beta$ emission lines, enforcing a shared narrow component, motivated by the physical expectation that the narrow emission arises from the narrow-line region. To model the compact origin of the broad emission in LRDs, we do not implement a typical broad gaussian component, which would be characterized by velocity widths ranging from 1000 - 10,000 km s$^{-1}$, instead, we model the broad component as an intermediate gaussian convolved with a symmetric exponential, as LRDs have been shown to feature exponential dominated broad profiles, resulting from electron scattering in dense gas \citet{Rusakov2026}, or a stratified broad line region \citet{Madau2026}. Symmetric exponential Balmer profiles have previously been observed in local Seyferts (see \citep[e.g.,][]{Laor2006, Kollatschny2013}. Similar to the model implemented in \citet{Matthee2026}, the intermediate gaussian depicts the intrinsic line kinematics, which is then scattered in dense gas to result in the exponential profile. To model the scattering fraction, we include a Thompson scattering parameter, $\tau_e$. 

For the narrow component, we vary the velocity width between 0 and 600 km s$^{-1}$, consistent with the expected narrow line region emission line widths in star forming galaxies \citep{Dai_2008}. We allow the narrow velocity to approach zero, as the convolution with the instrumental line velocity allows us to assess whether the narrow component is spectrally resolved. The narrow Balmer emission width is further constrained by enforcing a shared velocity width with the forbidden emission lines [\ion{O}{3}]$\lambda$4959, [\ion{O}{3}]$\lambda$5007, the [\ion{N}{2}] doublet, and the [\ion{S}{2}] doublet. For the intermediate gaussian, we vary the velocity width from 10 to 2000 km s$^{-1}$, and we vary the exponential width from 100 to 3000 km s$^{-1}$. We allow $\tau_e$ to vary between 0 and 1, representing a fully unscattered gaussian and complete scattering respectively. In general, our model setup is similar to the one implemented in \citet{Matthee2026}, with two key differences: we tie the line centers to the systemic redshift, and we do not assume a complete covering fraction for the narrow Balmer absorption.

To model the Balmer absorption feature that is often observed in both H$\alpha$ and H$\beta$ lines, we include a shared absorber applied to the broad component of both Balmer emission lines following the form from \citet{juodzbalis_24}:
\begin{equation}
    F_{obs} = F_{o} * [ 1 + C_f + C_f * e^{-\tau}]
\end{equation}
where
\begin{equation}
    \tau_\lambda = \tau_0 * exp \left[ \frac{0.5 * (\lambda - \lambda_{0,abs})^2}{\sigma_{abs}^2} \right]
\end{equation}
 where $\tau$ is the optical depth of the absorbing screen, defined by a gaussian line profile, and $C_f$ is the covering fraction of the absorber. We decouple the optical depth of the two Balmer lines by fitting both absorption Gaussian with independent widths and amplitudes, consistent with previous results that show that the Balmer absorption in LRDs is not always consistent with the expected $\tau_0$ ratio based on the bound-bound transition strengths for neutral hydrogen \citep{deugenio_2025_blackthunder, lambrides_discovery_2025}. And lastly, to ensure robust absorption detection, we fit all our objects with both the absorbed and unabsorbed models, and utilize a BIC criterion of 10 to classify them. This step acts as a safeguard for cases where the model erroneously predicts undetected absorption.

\begin{figure*}[!t]
    \centering
    \includegraphics[width=\linewidth]{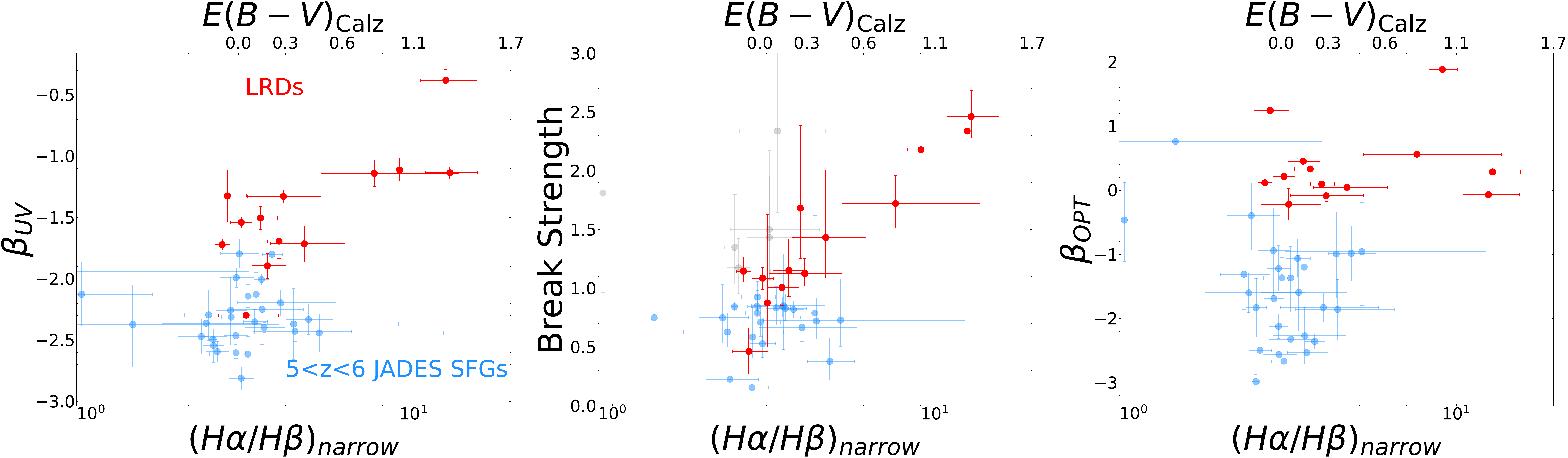}
    \caption{Narrow-line Balmer decrement plotted against (left) the observed UV continuum slope, $\buv$, (middle) the Balmer break strength, and (right) the rest-frame optical continuum slope, $\bopt$ for the 13 LRDs in our sample. \citet{Calzetti_2000} derived $E(B-V)$ values are indicated on the top axis of each panel. Red points indicate measurements for the LRDs, while blue points represent measurements for the $5<z<6$ JADES SFGs. In the middle panel, grey points indicate SFGs for which absorption features within the two wavelength bins used to measure the break strength results in an overestimation of the break strength; the true break strength for these objects would otherwise be less than 1. The narrow-line Balmer decrement correlates with $\buv$, consistent with the interpretation that the UV continuum slope of some LRDs is reddened by large-scale dust attenuation. The narrow-line Balmer decrement is also correlated with the Balmer break strength, while the $\bopt$ shows no systematic trend with the narrow-line Balmer decrement, perhaps indicative of a distinct physical origins for the rest frame UV and optical SED components in LRDs.}
    \label{fig:bdec}
\end{figure*}

\subsection{SED Fitting}

In order to place constraints on the stellar emission from the host galaxy, we use Prospector \citep{Johnson2021} with the Flexible Stellar Population Synthesis (FSPS: \citealt{Conroy2009, Conroy2010}) via python-FSPS \citep{ForemanMackey2014}, assuming a \citet{Chabrier_2003} initial mass function. We adopt a simple parametric, delayed $\tau$ star formation history (SFR $\propto t,e^{-t/\tau}$) with the age left free and the $e$-folding time fixed to 1 Gyr to ensure the model is star-forming and not quiescent. We fix the stellar metallicity to $\log(Z_\star/Z_\odot) = -1$ \citep{Nikopoulos2026}. Nebular emission is included self consistently from the CLOUDY grids of \citet{Byler2017} with the gas-phase metallicity tied to the stellar value and the ionization parameter $\log U$ is left free. Dust attenuation is modeled as a single starburst curve applied as a uniform foreground screen to the entire stellar population and to the nebular emission; we do not include a separate birth-cloud component, so young and old stars experience the same diffuse optical depth  $\hat{\tau}_V$ (equivalently $E(B{-}V) = 1.086,\hat{\tau}_V / R_V$), which we leave free. We fix each source to its spectroscopic redshift and apply intergalactic-medium attenuation. The spectrum is fit jointly with the NIRCam photometry; broad Balmer emission-line regions including the \ion{O}{3} emission lines are masked throughout. We sample the posterior with the dynesty nested sampler \citep{2020MNRAS.493.3132S}.

We perform the SED fitting analysis in two stages. In Stage 1 we isolate the rest-frame ultraviolet continuum, restricting the fit to rest-frame wavelengths $\lambda_{\rm rest} < 4200$ \AA. We extend the UV fit beyond the Balmer limit to explore whether the stellar template alone is able to account for the Balmer break. Additionally, this extension of the fitted spectrum beyond the Balmer limit allows for the nebular emission lines between the Balmer limit and 4200 \AA to constrain the ionization parameter. The output yields the stellar mass, age, $E(B{-}V)_{stellar}$, and $\log U$, which are determined primarily from the UV continuum. This fit procedure provides the cleanest test of whether reddened stellar UV emission can reproduce the observed UV continuum and its slope. 

In Stage 2 we test whether that same stellar solution, together with an additional rest-optical component, can reproduce the full spectral energy distribution. We freeze all stellar and nebular parameters at their Stage-1 posterior medians and add an empirical modified blackbody to represent the characteristic red optical excess of LRDs (similar to \citealt{DeGraff_2025}), and vary its amplitude $A$, power-law modifier $\beta$, and peak wavelength. This component carries no emission lines, so the predicted nebular emission is identical in both stages. We deliberately keep this optical component agnostic as to its physical origin, using it only to model the rest-optical continuum so that the result of the stellar UV fit from Stage 1 is not affected by the optical data.

For the SFGs, we perform the fit in a single stage using the same star-forming template as the LRDs. The primary differences from the LRD fitting procedure is that we do not include an additional rest-optical modified BB component, and we allow the mettallicity to vary freely, as the unmasked emission line regions provide sufficient constraints.

\section{Results}
\label{sec:result}

In this section, we first report on the Gaussian fitting results and compare them with existing LRD analyses based on medium-resolution spectroscopy. Then, we use the narrow-line Balmer decrement to constrain the large scale attenuation in LRDs. Finally, we study how large scale dust attenuation affects the spectral shape of LRDs.

\subsection{Line fitting result}
\label{sec:fit}

 Figure \ref{fig:cutout} shows an example of our fitting result for a RUBIES 42046 at z = 5.28. We have indicated the broad and narrow components of $\Ha$ and $\Hb$ alongside the strong [\ion{O}{3}] doublet, the [\ion{N}{2}] doublet, and the continuum dominated [\ion{S}{2}] doublet. The rest of the fitting results for the other 12 sources are presented in Appendix \ref{sec:appendix}. Overall, our fitting results suggest that both Balmer lines are well described by a shared broad and narrow component. We note, however, that the broad and narrow component likely arise from distinct physical regions within the LRD, a point we return to in Section \ref{sec:conclusions}.

For all 13 LRDs in our medium-resolution subsample, we successfully decompose the H$\alpha$ emission line into the intermediate and exponential wing components. The broad component of the $\Hb$ emission line is often undetected, likely a result of the nuclear obscuration in LRDs. It has been suggested the broad Balmer emission could be subject to collisional excitation in a high density environment, characterized by the exponential profile of the Balmer emission \citep{Rusakov2026}, causing the broad line Balmer decrement to deviate from case B recombination, and resulting in excessively large Balmer decrement values. In most of the 13 LRDs, the intrinsic velocity width of the narrow emission is $\lesssim$ 20 km $s^{-1}$, suggesting that the narrow line is unresolved even in R $\sim$ 1000 medium-resolution spectra. However, we resolve the narrow line width in five of the 13 LRDs; JADES-GDN 68797 ($\sigma_n$ $\simeq$ 80 km $s^{-1}$ ), EGS 51623 ($\sigma_n$ $\simeq$ 210 km $s^{-1}$), RUBIES 920396 ($\sigma_n$ $\simeq$ 90 km $s^{-1}$), RUBIES 42046 ($\sigma_n$ $\simeq$ 90 km $s^{-1}$), RUBIES 47509 ($\sigma_n$ $\simeq$ 70 km $s^{-1}$), which are consistent with previous narrow-line width measurements of LRDs using high-resolution R $\sim$ 2700 IFU spectroscopy \citep{Ji_2025}. 

It has been suggested that the detection of a narrow Balmer absorption in LRDs is likely tied to the spectral resolution \citep{Shangguan2026}, with some publications claiming to have detected unresolved absorption through line fitting \citep{Nikopoulos2025}. To ensure we do not miss any undetected absorption in our analysis, we fit all our objects with both absorbed and unabsorbed components. Despite cases where the absorption model suggested there was unresolved absorption, we only find three LRDs with robust ($\Delta$BIC $>$ 10) narrow Balmer absorption components, these coincided with the three LRDs we visually classified as having narrow Balmer absorption. In the three LRDs with a narrow Balmer absorption component, the optical depth ratio deviates from the expected ratio of $\tau_{0,H\alpha}$ / $\tau_{0,H\beta}$ $\sim$ 7.2 based on the oscillator strength, with the $\Hb$ optical depth being much larger than expected, suggesting that the Balmer absorption is inconsistent with a simple screen absorber geometry. However, the absorption does appear to share a common origin, as the shared velocity offset provides an adequate fit for all three LRDs. We present the rest of the fitting results in the Appendix.

\subsection{Narrow line Balmer Decrement in Little Red Dots}
\label{sec:narrow_line}

In this section, we analyze the narrow-line decrement as a proxy for large scale dust obscuration in LRDs. Previous narrow-line decrement measurements of LRDs have reported a wide range of values, painting an inconsistent picture of the large scale dust attenuation present in these systems. \citet{killi_24_lrd_spectrum} reports a narrow-line Balmer decrement value of 4 for a LRD at $\sim$ 4.53, \citet{de_graff_25a} reports a value of  $\sim$ 7.4 for \textit{The Cliff}, and \citet{Deugenio_25} reports a narrow-line decrement of only $\sim$ 3.4 for \textit{Irony}, a RUBIES LRD at z = 6.68. These analyses have employed both low-resolution PRISM \citet{killi_24_lrd_spectrum} and medium-resolution \citet{de_graff_25a, Deugenio_25}, alongside different line decomposition techniques, \citet{killi_24_lrd_spectrum} employs a double gaussian component for fitting the Balmer emission, \citet{de_graff_25a} adopts a Lorentzian profile, while \citet{Deugenio_25} implements a Gaussian convolved with an exponential. The range of line fitting approaches reflects the ambiguous nature of the line emission in LRDs, making it difficult to compare narrow decrement values across the literature, although there is now consensus that the broad Balmer lines in LRDs are consistent with exponential profiles \citep{Rusakov2026}. With our larger sample of LRDs with medium-resolution spectroscopy, and a better understanding of how to model the broad emission in LRDs, we are therefore well-positioned to investigate what the narrow-line decrement can reveal about the physical nature of LRDs.

Figure \ref{fig:bdec} shows the measured narrow-line Balmer decrement alongside the PRISM-derived $\buv$ (left), break strength (middle), and $\bopt$ (right) for the 13 LRDs. We also plot the $\buv$ and $\bopt$ of JADES z $\sim$ 5-6 SFGs for comparison to the LRDs. We treat the PRISM-derived Balmer decrements of the SFGs as narrow-line decrements, since the Balmer emission lines can be characterized by a single gaussian. The narrow-line decrement values for the SFGs are generally close to the case B recombination value of 2.86, which is unsurprising given the bulk of the star forming galaxies at z $\sim$ 5-6 are expected to not have much dust. The narrow-line decrement values for the LRDs span a larger range of values; while most of the 13 LRDs have decrement values $\sim$ 3.0, indicative of minimal large scale dust attenuation, we do find a subset with elevated narrow-line Balmer decrement ($\gtrsim$ 7) corresponding to an $E(B-V)$ $\sim$ 0.7 - 1.4 assuming the \citet{Calzetti_2000} attenuation law, indicating moderate to significant large scale dust in these LRDs. One of the LRDs analyzed here, JADES-GN-73488, was previously reported by \citet{Nikopoulos2025} as having a narrow Balmer decrement of 1.88, thus they argue it violates case B recombination. However, from our line fitting analysis, we recover a narrow Balmer decrement of 2.588, a value that is in better agreement with case B recombination.

In the following sections, we will explore how the presence of large scale dust attenuation affects different parts of the LRD spectroscopy. 

\subsubsection{UV Continuum}

We begin by examining the relationship between the narrow-line Balmer decrement and the UV continuum slope, $\buv$. As shown in the left panel of Figure \ref{fig:bdec}, the JADES 5 $<z<$ 6 SFGs exhibit blue UV emission (-2.8 $\lesssim \buv \lesssim$ -1.8); in comparison, LRDs exhibit systematically redder UV emission, spanning a broader range of -2.5 $\lesssim \buv \lesssim$ -0.3. We find only one LRD, JADES-GDN-39353, with $\buv$ comparable to the SFG population. We identify a broad correlation between $\buv$ and the narrow-line Balmer decrement in LRDs, whereby LRDs with high narrow-line Balmer decrements also exhibit red UV slopes. 

The relationship between the Balmer decrement and the $\buv$ is well established for local star forming galaxies \citep{Calzetti_94_uv_slope,Calzetti_2000, Reddy2015}, where the correlation is expected to arise from the reddening effect of large scale dust attenuation on the UV continuum. If the intrinsic UV continuum of LRDs is blue (i.e., $\buv$ is more negative than observed), as would be expected if the UV emission arises from a host galaxy, then large-scale dust attenuation reddens the UV continuum, giving rise to the observed correlation shown in the left panel of Figure \ref{fig:bdec}. This scenario has already been demonstrated through simulations, to adequately reproduce the rest UV photometric selection criteria used to identify LRDs in JWST fields \citep{Volonteri2025}. Under the assumption that the UV emission arises from a dust-attenuated host galaxy, our median E(B-V) is 0.19, which when multiplied by $R_v$ = 4.05 \citep{Calzetti_2000}, returns a median $A_{\rm V}$ $\sim$ 0.8, in agreement with the median $A_{\rm V}$ $\sim$ 0.8 predicted by \citealt{Volonteri2025}.

\begin{figure*}
    \centering
    \includegraphics[width=\textwidth]{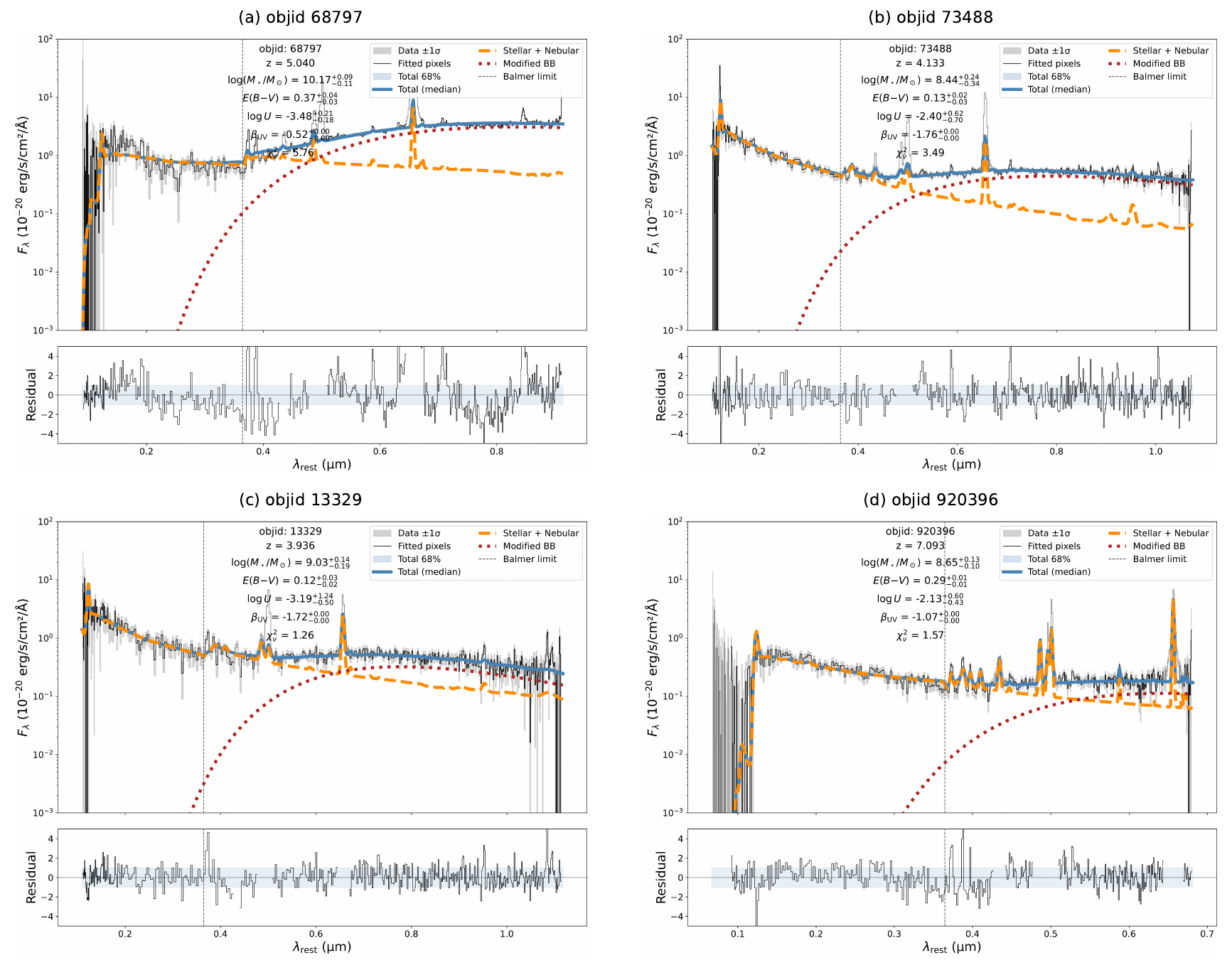}
    \caption{Prospector SED fit result for four of the LRDs included in our analysis. A.) Top panel: Prism spectrum of JADES-GDN-68797 over-plotted with the SED fit model (blue) alongside the templates ( stellar (orange) and Modified black body ( burgundy)) that are used in the modeling. The text label  has been added to highlight key model parameters: stellar mass, stellar attenuation, ionization parameter, and the UV slope of the SED fit model. We have also included the reduced chi squared to quantify the good ness of our fit. Bottom panel: The residual ( data - model). The masked regions (broad Balmer emission and \ion{O}{3} can be easily identified in the residual plot as blank sections. B.) Same as A but for JADES-GDN-73488. C.) Same as A but for JADES-GDS-13329. D.) Same as A but for MOM-UDS-920396.}
    \label{fig:spectrum-fit}
\end{figure*}

\subsubsection{Break Strength}

To investigate how large-scale dust attenuation affects the ``V'' shaped SED of LRDs, we examine the relationship between the narrow-line Balmer decrement and the Balmer break strength. We measure the break strength following the method of \citet{de_graff_25a}, defined as the the average flux density ratio between two rest-frame wavelength bands: [0.362,0.372] $\mu$m and [0.40,0.41] $\mu$m, with uncertainties estimated using 500 random Gaussian draws from the error spectrum. We measure the break strength for both the SFG and LRD population and plot them against their respective narrow line decrements in the middle panel of Figure \ref{fig:bdec}.

The SFGs exhibit break strength values $<$ 1, consistent with the expected power-law continuum shapes of young, relatively dust-free star-forming galaxies. A small number of exceptions are present, including one object with an anomalously high break strength $\sim$ 2.4. Upon further inspection, we discover the reason for the high SFG break strength measurements to be an absorption feature present in the continuum within the two wavelength bins used to measure the break strength, rather than a genuine incidence of a high Balmer break.

All LRDs in our sample exhibit break strength values $\gtrsim$ 1, with the exception of JADES-GDN-39353, which has a break strength of $\sim$ 0.5. This outlier LRD, similarly to its $\buv$ value, is consistent with the break strength values of the SFG population. We find no evidence of continuum absorption-inflated break strength values in the LRD population, and a clear separation in break strength is apparent between the two populations. We also identify a strong correlation between the break strength and the narrow-line Balmer decrement, such that LRDs with larger narrow-line Balmer decrement values systematically exhibit stronger breaks. Unlike the case for $\buv$, no established relationship between the Balmer break strength and the Balmer decrement is seen for star-forming galaxies. While the range of break strength values explored here, [0.5, 2.5], is consistent with the theoretical range achievable in star-forming galaxies \citep{Forrest2018}, it remains unclear whether the observed correlation reflects large-scale dust attenuation strengthening the Balmer break, or whether it is instead a secondary consequence of the correlation between $\buv$ and the narrow-line Balmer decrement. We discuss this further in Section \ref{sec:conclusions}.

We further find that dividing the population by break strength into strong breakers ($>$ 1) and power-law SEDs ($<$ 1) following the classification scheme of \citealt{Barro_25}, the transition from the power-law SED to a strong breaker coincides with increasing narrow line decrement. This result is in tension with the scenario 2 explanation of the diversity in LRD spectra proposed in \citet{Barro_25}, in which the transition is attributed to nuclear attenuation within an otherwise unattenuated host galaxy. In contrast, our result indicates that the transition is associated with large scale attenuation of the host galaxy. This interpretation remains agnostic as to whether the nuclear engine of LRDs is consistent with a black hole star or a conventional broad line AGN.

\subsubsection{Rest Optical Continuum}

The approach adopted here, in which we treat the break strength and the rest-frame optical continuum separately, implicitly allows for a scenario where the two wavelength components may each have distinct physical origins. This stands in contrast to previous analyses of LRDs, which have argued that the rest-frame optical continuum, including the Balmer break, can be well-described by a single blackbody template, attributed to thermal emission from the central enshrouded black hole under the black hole star model \citep{DeGraff_2025}. In that framework, the rest-frame optical continuum and the Balmer break are assumed to share a common physical origin. We test this assumption directly by examining whether dust attenuation affects the observed Balmer break and rest-frame optical continua similarly, by comparing the narrow-line Balmer decrement with the rest-frame optical continuum slope, $\bopt$.

As shown in the right panel of Figure \ref{fig:bdec}, there is clear separation between the $\bopt$ values for the SFGs and the LRDs. LRDs exhibit either flat ($\bopt$ $\simeq$ 0) or positive $(\bopt > 0$) rest optical continuum slopes compared to SFGs which exhibit primarily negative slopes ($\bopt < 0$). This behavior reflects the selection criteria for LRDs, as LRDs are selected based on their non-negative optical slopes. However, unlike the $\buv$ and break strength, we find that the $\bopt$ for LRDs exhibits no systematic trend with the narrow-line Balmer decrement, appearing essentially scattered across the full range of decrement values.

\begin{figure}
    \centering
    \includegraphics[width=\linewidth]{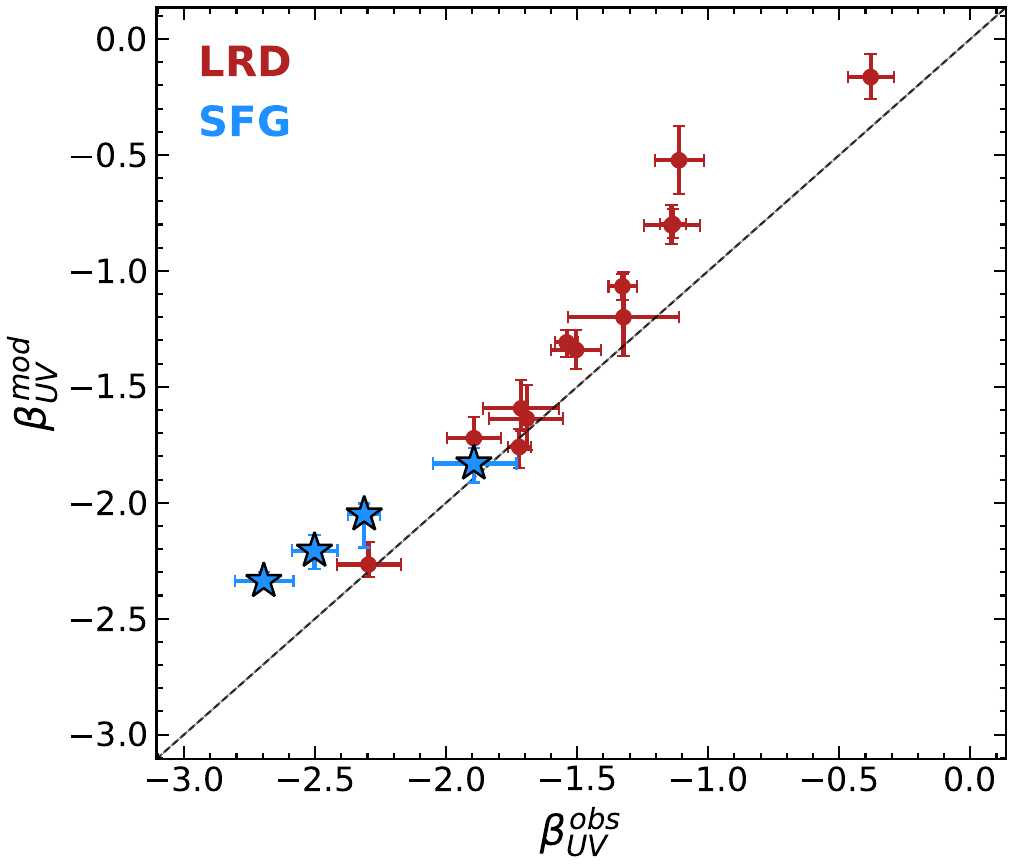}
      \caption{Comparison of the UV slope between the measured value from the PRISM spectroscopy and the value predicted by the SED fit model. The red points represent the thirteen LRDs with medium-resolution spectroscopy, while the blue represent the composite SFGs. In general, there is good agreement between the model and the data for LRDs with bluer slopes ( $\buv$ $<$ -1.5), while the model predicts bluer UV slopes than observed for LRDs with redder $\buv$ values.}
    \label{fig:compare_sed_observed}
\end{figure}

\begin{figure}
    \centering
    \includegraphics[width=\linewidth]{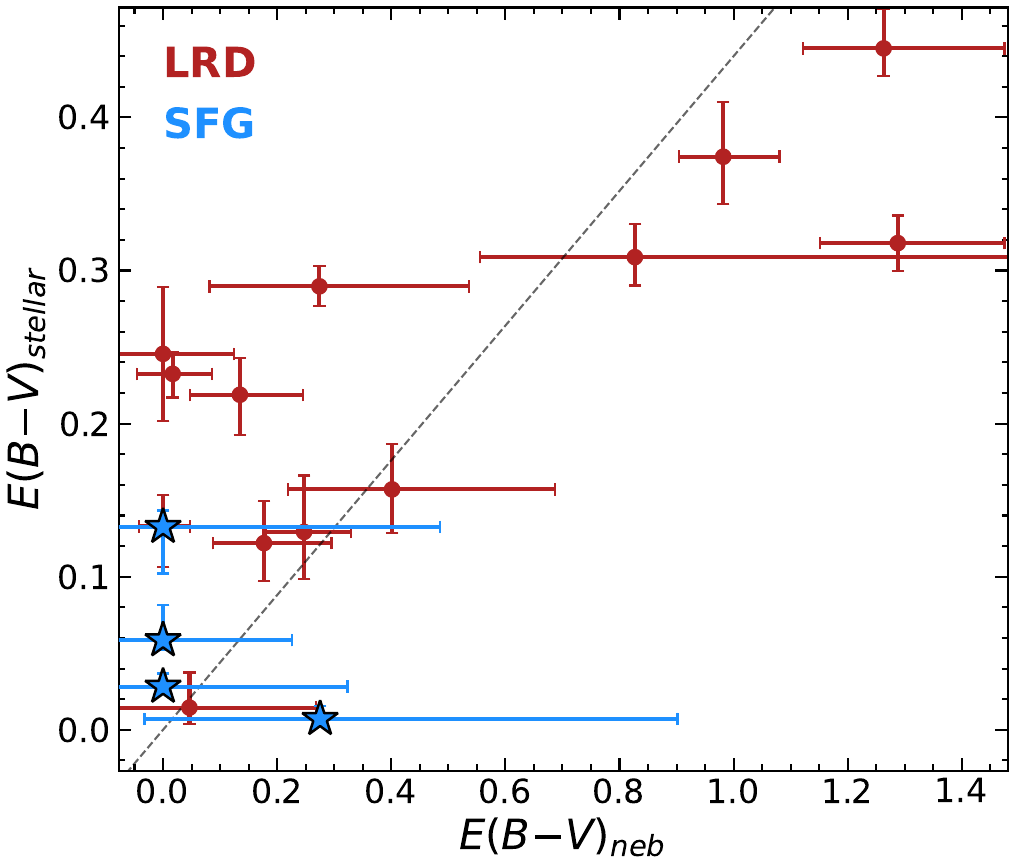}
      \caption{Comparison of the stellar attenuation from the SED fitting results with the nebular attenuation inferred from the narrow-line Balmer decrement. The red points represent the thirteen LRDs with medium-resolution spectroscopy, while the blue points represent the composite SFGs. The dashed line represents the expected relationship between the two parameters from the  \citet{Calzetti_2000} attenuation curve. There is broad agreement between the distribution for the LRDs and the expected relation; however, a significant subset exists for which the model favors stellar attenuation while the narrow-line decrement predicts negligible attenuation. The SFGs in general do not lie on the relation, with three of the stacked points lying above the line ( more stellar attenuation than predicted from the Balmer decrement) and one below. We attribute this scatter to the PRISM resolution rather than a genuine mismatch between the nebular and stellar attenuation in SFGs. }
    \label{fig:compare_ebv_sed_observed}
\end{figure}

\subsection{SED fitting of LRD UV continua}

Based on the broad trend identified between the $\buv$ and the narrow-line Balmer decrement in the left panel of Figure \ref{fig:bdec}, we argue that the UV continuum in LRDs originates from large scale dust attenuated stellar UV emission. However, it remains unclear whether the UV emission in LRDs can be attributed entirely to the host galaxy, or whether an additional source of emission must be accounted for to properly describe the UV emission in LRDs. Additionally, under the assumption that LRDs reside in host galaxies, it is worth examining what intrinsic differences exist between the LRD host galaxies and SFGs at similar redshifts. This distinction is especially important given the distribution observed in the left panel of Figure \ref{fig:bdec}, where a sub-population of the LRDs with narrow line Balmer decrement values similar to those of the SFGs still exhibit redder $\buv$ values. 

We therefore perform SED fitting analysis of the PRISM spectra of the 13 LRDs analyzed above to examine if a stellar template could provide an adequate fit to the UV continuum in LRDs. We aim to constrain the intrinsic properties of the stellar templates needed to model the UV continuum of LRDs and how these compare to those of the SFGs. Figure \ref{fig:spectrum-fit} shows an example of the full composite fit for four LRDs representative of the diversity of the thirteen LRDs analyzed: JADES-GDN-68797, JADES-GDN-73488, JADES-GDS-13329, and MOM-UDS-920396. The best-fit results of each stage is indicated on each sub-figure; the stellar and nebular emission component (orange) from the stage 1 fit, and the modified BB (burgundy) from the stage 2 fit. We have also labeled each sub-figure with the best-fit parameter values from the SED fitting analysis: including the stellar mass, the stellar attenuation, the ionization parameter and the UV slope. We then extract and compare these key parameters from the SED fitting from across the full sample of SFGs and LRDs. The results of this comparison analysis are summarized below.

\begin{figure*}[t]
    \centering
    \includegraphics[width=\linewidth]{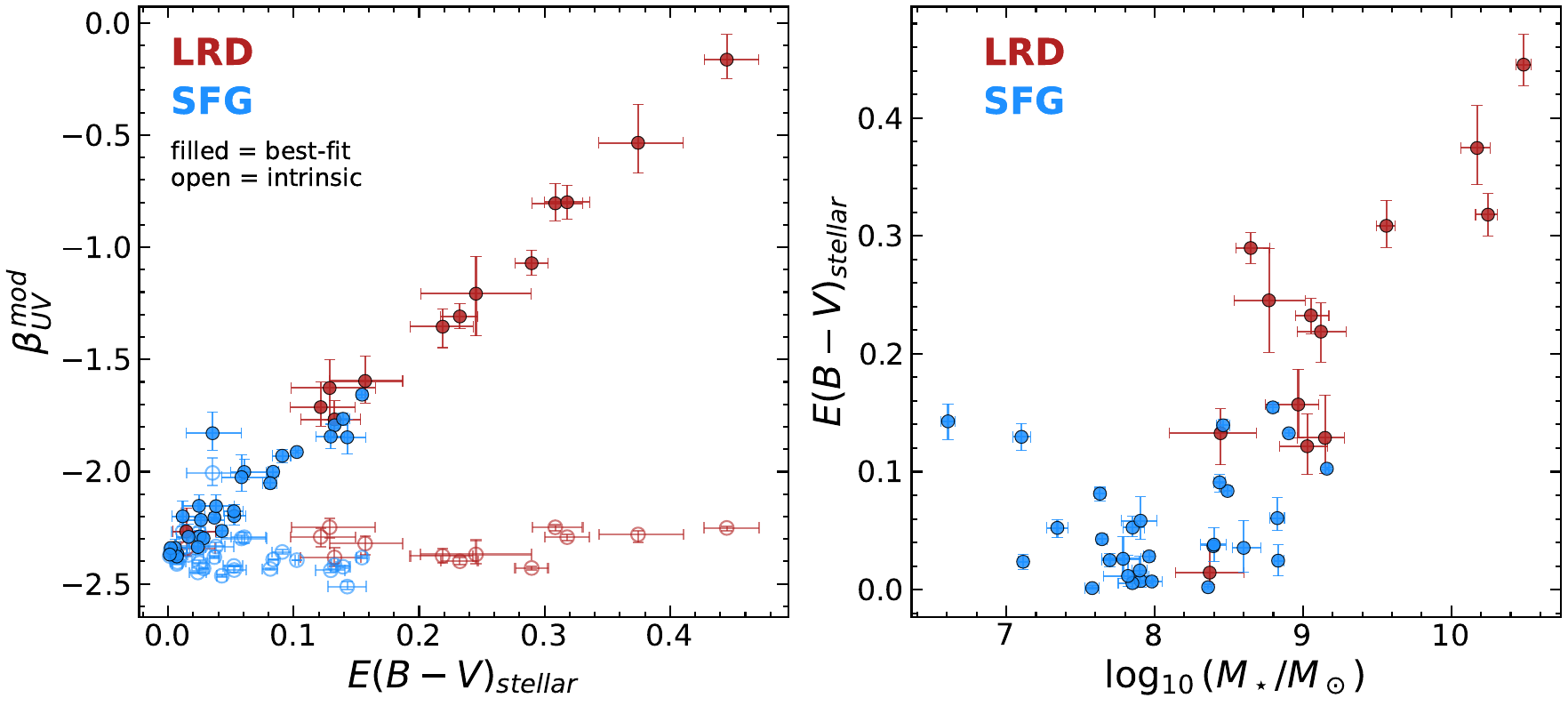}
    \caption{Best-fit results from the SED fitting analysis. Left panel: the model UV slope, $\buvmod$, plotted alongside the stellar attenuation, $E(B-V)_{stellar}$. The red points represent the LRDs, while the blue points represent the individual SFGs. The filled circle show the best-fit result, while the open circle represents the intrinsic UV slopes of the stellar templates used for the SED fit. $\buvmod$ is tightly correlated with the $E(B-V)_{stellar}$, and the two population occupy distinct regions: LRDs exhibit higher $E(B-V)_{stellar}$ values and redder $\buvmod$, while the opposite holds for the SFGs. Right panel: $E(B-V)_{stellar}$ plotted against stellar mass. The color scheme is the same as the left panel. Overall, the best-fit stellar masses for the LRDs are typically higher than those of the SFGs. $E(B-V)_{stellar}$ remains constant and negligible up to a stellar mass of  $\log(\Mstar/M_{\odot})$ $\sim$ 8.2, beyond which a clear correlation emerges between the two parameters. Overall, these two panels present a picture in which the LRD host galaxies are dustier than SFGs, with the excess dust appearing tied to the mass difference between the two populations. }
    \label{fig:mass_v_dust}
\end{figure*}

\subsubsection{UV slopes and stellar attenuation of the host galaxy template}

Figure \ref{fig:compare_sed_observed} shows the the UV slope derived from the SED fitting, $\buvmod$, alongside the the observed $\buv$ values, $\buvobs$, for the 13 LRDs with medium-resolution spectra. Due to the SFGs being less luminous than the LRDs and the comparatively worse resolution of the PRISM spectroscopy to the medium-resolution spectroscopy, the Balmer decrement measurement is noisier for the SFGs than the LRDs. To compare the SFGs to the LRDs with comparable signal-to-noise, we bin the SFGs based on their $\buvmod$ values and then stack them to generate composite values, with the results represented as blue stars. We have also added a dashed line to represent equality between the two parameters. We find that for LRDs with the bluest UV slopes ($\buvobs$ $<$ -1.5), the $\buvmod$ is in agreement with the $\buvobs$. This agreement suggests that for the LRDs with the bluest UV slopes, it is possible and indeed likely that the UV emission arises primarily from a host galaxy. However for LRDs with redder $\buvobs$ values, ($\buvobs$ $\geq$ -1.5), the stellar template predicts bluer $\buvmod$, which can be observed in Figure \ref{fig:compare_sed_observed} as the red data points raised upwards away from the line of equality. For the SFGs, we find a general agreement between $\buvmod$ and $\buvobs$.

Figure \ref{fig:compare_ebv_sed_observed} shows the the dust attenuation required for the stellar template to fit the observed UV continuum, $E(B-V)_{stellar}$, alongside the nebular attenuation calculated from the narrow-line Balmer decrement the \citet{Calzetti_2000} attenuation law, $E(B-V)_{neb}$. We adopt the \citet{Calzetti_2000} attenuation curve throughout our analysis, therefore, we have added a line reflecting the expected relationship between the stellar attenuation and the nebular attenuation: 
\begin{equation}
    E(B-V)_{stellar} = 0.44 \times E(B-V)_{neb} 
\end{equation}

Under the assumption that the narrow line emission in the medium-resolution arises solely from a host galaxy component and that the dust attenuation in LRDs can be described using the \citet{Calzetti_2000} curve, LRDs for which the stellar template accurately fits should lie on the dashed line. However, the results in Figure \ref{fig:compare_ebv_sed_observed} reveal a more complicated picture of the origin of the narrow line emission. While the relationship between the nebular attenuation and the stellar attenuation of the LRDs broadly follows the \citet{Calzetti_2000} relation, there are some clear outliers -- in particular, LRDs in the lower left corner of the figure, where the narrow line Balmer decrement implies negligible nebular attenuation, yet the SED fit unambiguously prefers significant stellar attenuation for the stellar template. Comparatively, we find that the stacked SFGs appear consistent with the expected relation within the 1 sigma error.

\begin{figure*}[!ht]
    \centering
    \includegraphics[width=\linewidth]{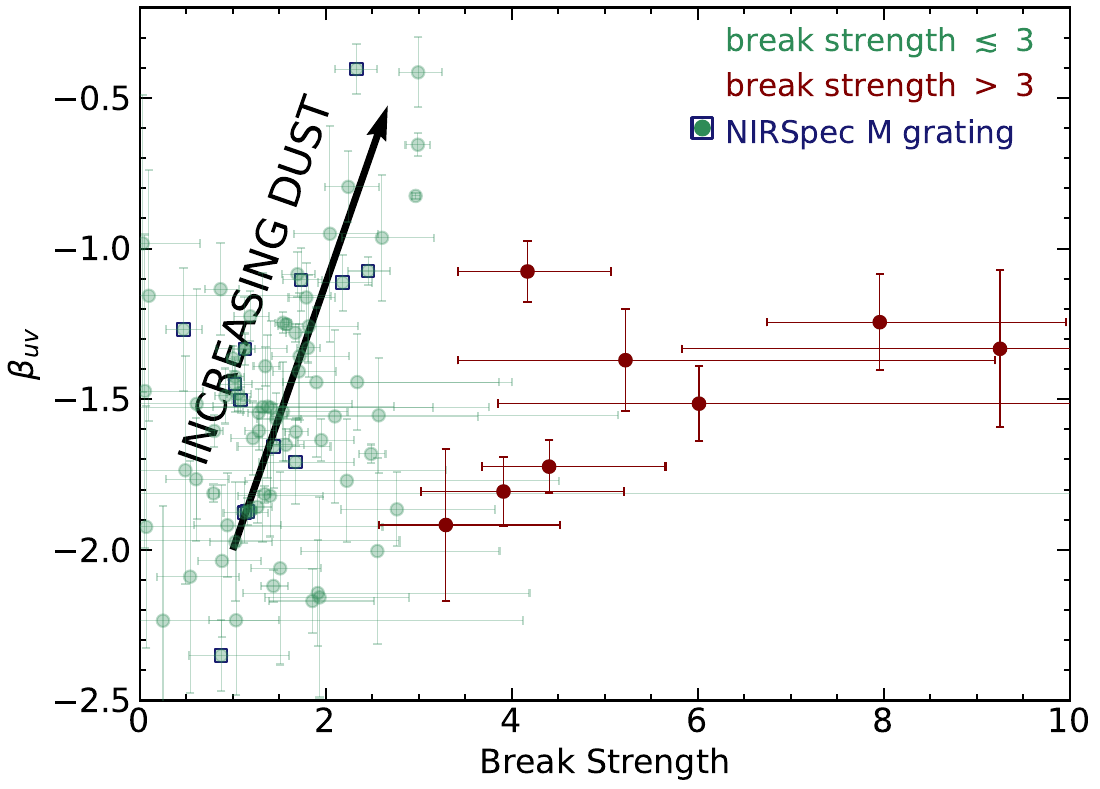}
    \caption{Balmer break strengths and observed UV continuum slopes ($\buv$) of the 89 NIRSpec PRISM LRDs in our sample. Green points represent LRDs with break strengths $\lesssim$ 3, and maroon points represent LRDs with break strengths $\gtrsim$ 3. The blue open squares represent the 13 LRDs with medium-resolution spectra. The black arrow indicates the direction of increasing large-scale dust attenuation, derived from the correlated evolution of $\buv$ and the Balmer break strength with increasing narrow-line decrement, as demonstrated in Figure \ref{fig:bdec}.} 
    \label{fig:uv_plus_break}
    
\end{figure*}
\subsubsection{LRD host galaxies and SFGs}

Figure \ref{fig:mass_v_dust} shows the distribution of three key parameters from the SED fitting analysis: the UV slope, $\buv$, stellar mass, $\Mstar$, and the stellar attenuation, $E(B-V)_{stellar}$. The left panel shows $\buvmod$ plotted against $E(B-V)_{stellar}$, essentially recreating the left panel of \ref{fig:bdec} with parameters derived from the SED fit. The intrinsic UV slopes of the stellar template used are indicated by open circles, with colors matched to the filled circles for each population. The tight relationship between $\buvmod$ and $E(B-V)_{stellar}$ indicates that the primary driver of the separation between the model UV slope for LRDs and SFGs is the amount of stellar attenuation applied to the UV continuum, a result that is further strengthened by the fact that the intrinsic templates for both populations are the same, as indicated by the range of intrinsic UV slopes.

The right panel shows the $E(B-V)_{stellar}$ plotted alongside the stellar mass. Here we aim to understand what intrinsic differences emerge between the models for the two populations, and whether the difference can account for the higher level of attenuation observed in the LRDs. We find the stellar models for the LRDs are more massive on average than those of the SFGs, with inferred stellar masses spanning 6.5 $\lesssim$ $\log(\Mstar/M_{\odot})$ $\lesssim$ 9.3 for the SFGs and 8.3 $\lesssim$ $\log(\Mstar/M_{\odot})$ $\lesssim$ 10.5 for the LRD host galaxies. Additionally, we find a broad correlation between stellar mass and stellar attenuation: the stellar attenuation remains low and approximately constant for the models with stellar mass below $\log(\Mstar/M_{\odot})$ $\lesssim$ 8.5, and follows a log-linear relation for the models with $\log(\Mstar/M_{\odot})$ $>$ 8.5. Similar relationships between the stellar mass and dust attenuation in galaxies have been previously reported (see Figure 10 in \citet{Markov2025}). These results suggest that if LRDs do reside in host galaxies, these host galaxies are likely to be more massive than typical SFGs at similar redshifts, and that stellar mass is directly correlated with the stellar attenuation present in the SED.

\subsection{Dust attenuation and the spectral shape of LRD}
We have now presented two complementary diagnostics of large scale dust attenuation in LRDs: measurements of the narrow-line Balmer decrement, and SED fitting analysis showing that the UV continuum in LRDs is consistent with a dust-attenuated stellar template. We now seek to present a comprehensive picture of how large scale dust attenuation shapes the observed spectral shape of LRDs. To this end, we plot $\buv$ alongside the break strength in Figure \ref{fig:uv_plus_break} for the full sample of JWST PRISM LRDs. To further characterize the behavior of LRDs in Figure \ref{fig:uv_plus_break}, we find it necessary to divide the population into two break strength regimes: break strength $>$ 3 and and break strength $\lesssim$ 3. 

For the sub-population of LRDs with break strength $\lesssim$ 3, we find that the correlation between the $\buv$ and the break strength is consistent with increasing dust attenuation, whereby a stronger Balmer break is accompanied by a redder $\buv$. The arrow in Figure \ref{fig:uv_plus_break} illustrates how this correlation evolves with increasing dust attenuation, following from the behavior observed in the left and middle panels of Figure \ref{fig:bdec}. This behavior is unsurprising given the already established connection between the break strength, $\buv$, and large scale dust attenuation. We find that roughly two-thirds of the LRDs reside in the 1 $<$ break strength $\lesssim$ 3 regime, and therefore are consistent with this dust attenuation model. We note, however, that while the selection method applied in \citet{DeGraff_2025} results in a high purity sample, the selection is not uniform, particularly when considering photometrically selected LRDs. As such, the fraction reported here may change considering the full population of LRDs available in literature. Nevertheless, our results indicate that careful consideration must be given to dust reddening when modeling the underlying rest-frame optical emission.

In the break strength $>$ 3 regime, we find that as the break strength increases and the Balmer break becomes more prominent, the $\buv$ remains constant at a moderately red slope. This behavior diverges from the dust attenuation trend observed in the break strength $\lesssim$ 3 regime, where increasingly prominent Balmer breaks are accompanied by the redder $\buv$ values. Unfortunately, none of the LRDs in this regime have available medium-resolution spectroscopy in DJA, so we are unable to directly measure their narrow-line Balmer decrements. Consequently, while large scale dust attenuation may still be present in these high break strength LRDs, other physical mechanisms appear to play a more dominant role in setting the spectral shape. 

Given the divergence between the two break strength regimes, we introduce two distinct subsets of the LRD population: one in which dust attenuation plays a dominant role in setting the spectral shape and red colors of LRDs, and a second in which other physical processes likely dominate. In the following sections, we examine composite spectra across the three regimes to understand how these trends manifest in the observed spectra. 

\begin{figure*}[!t]
    \centering
    \includegraphics[width=\textwidth]{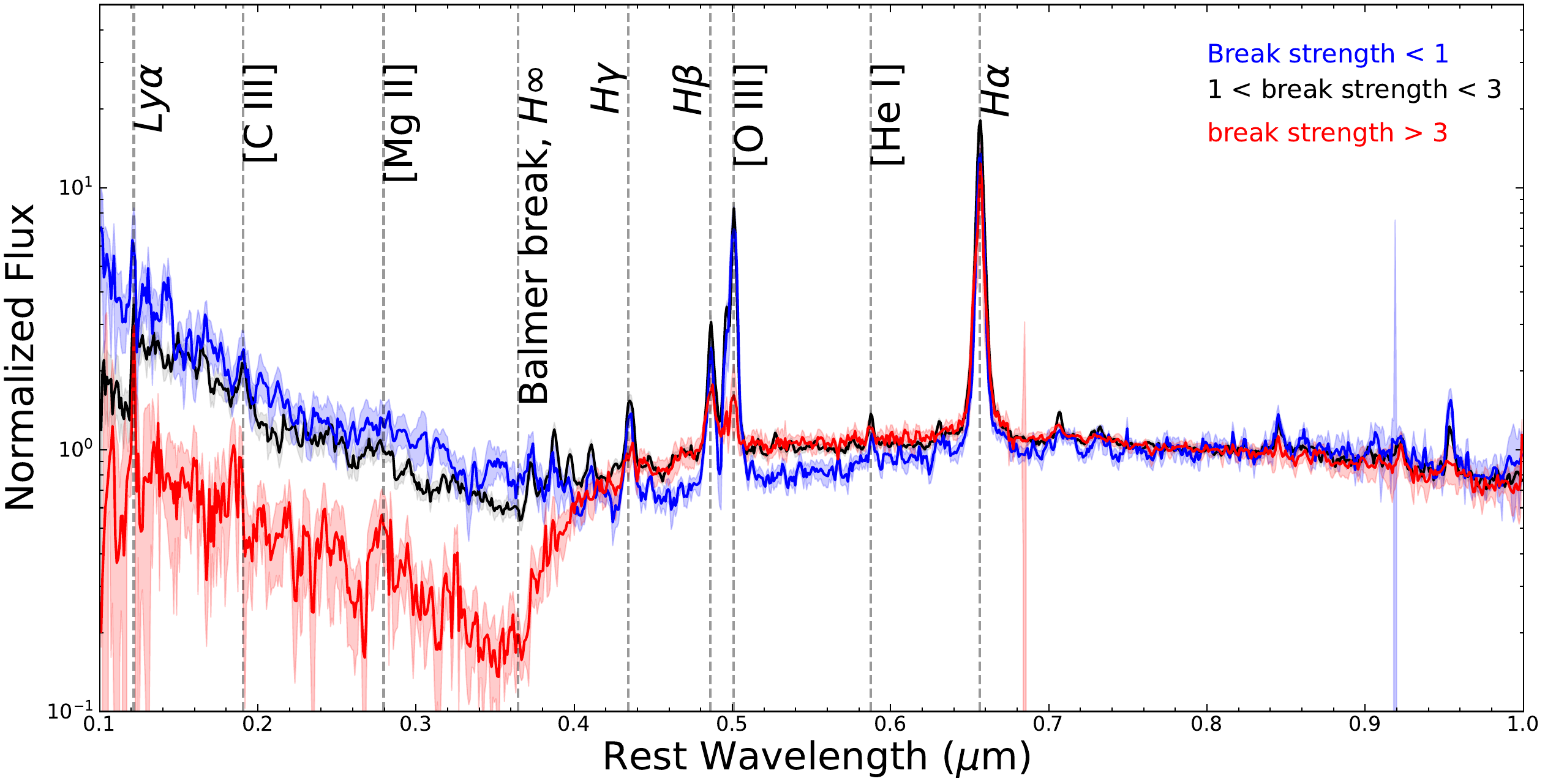}
    \caption{Composite JWST NIRSpec PRISM spectra for sub-populations of LRDs classified by Balmer break strength. The blue line represents the power-law (break strength $<$ 1) population, the black line represents the intermediate (1 $<$ break strength $\lesssim$ 3) population, and the red line indicates the extreme breaker (break strength $\gtrsim$ 3) population. The shaded region represents the bootstrap resampled estimated error for the composites. All spectra are normalized over the wavelength range 0.77 -- 0.83 $\mu$m. Vertical markers indicate the location of prominent emission lines.  }
    \label{fig:composite_all_three}
\end{figure*}

\begin{figure*}[!t]
    \centering
    \includegraphics[width=\textwidth]{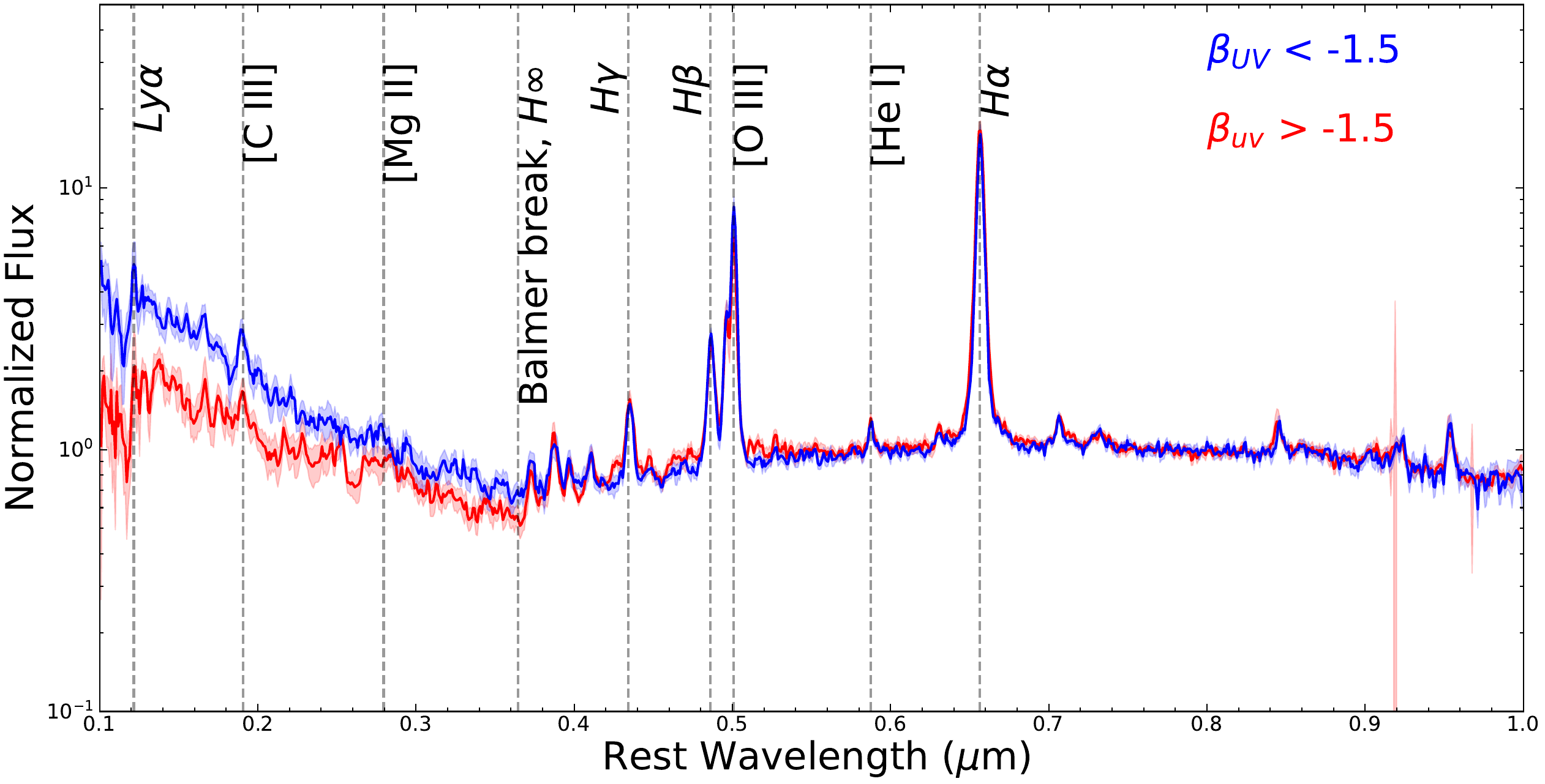}
    \caption{Composite spectra for LRDs with Balmer break strength $\lesssim$ 3, divided by $\buv$. The blue line represents the bluer UV slope ($\buv$ $<$ -1.5) population, and the red line represents the redder UV slope ($\buv$ $>$ -1.5) population. All spectra are normalized over the wavelength range 0.77 -- 0.83 $\mu$m. Vertical markers indicate the location of prominent emission lines.}
    \label{fig:composite_uv}
\end{figure*}

\subsubsection{Sub-populations of LRDs}

In the previous section, we discovered a naturally arising distinction among LRDs based on their break strength values. The emission from LRDs with break strength values $\lesssim$ 3 appear to be determined by different physical mechanisms than for those with break strengths $>$ 3. To characterize the typical spectral properties in the different break strength regimes, we stack the full JWST NIRSpec PRISM spectra of LRDs in the each sub-population. To build the composite spectra, we first build a rest-frame wavelength grid of 1000 steps between 0.1 and 1 micron. We then resample each spectrum onto this wavelength grid. Next, we stack the resampled spectra and extract the composite spectrum. We estimate errors by performing a 500 iteration bootstrap resampling of the stack, extracting the median for each bootstrap and then taking the standard deviation of those medians. In total, we classify 89 LRDs here, 20 in the break strength $<$ 1 regime, 61 in the 1 $<$ break strength $\lesssim$ 3, and 8 in the break strength $>$ 3 regime. The composite spectra are shown in Figure \ref{fig:composite_all_three}, and we explore the similarities and differences between each sub-population across the UV and optical wavelengths in the following section.

\subsubsection{UV Properties of Composites}

Here we explore the spectral properties of the different break strength regimes at UV wavelengths. We define the UV as everything blueward of the Balmer limit, H$\infty$. Overall, we find a relatively fainter UV continuum in the break strength $>$ 3 regime compared to the break strength $\lesssim$ 3 regime. Within the break strength $\lesssim$ 3 regime, the UV continuum is remarkably similar across sub-regimes. This is somewhat surprising given the correlation shown in Figures \ref{fig:bdec} and \ref{fig:uv_plus_break} which demonstrates that the UV slope reddens with increasing break strength. However, the bulk of the LRDs in the 1 $<$ break strength $\lesssim$ 3 regime have $\buv$ values similar to the break strength $<$ 1 regime, causing them to dominate the composite and resulting in similar composite UV slopes for the two sub-regimes. Nevertheless, we do observe a redder UV continuum slope in 1 $<$ break strength $\lesssim$ 3 regimes compared to the break strength $<$ 1 regime, consistent with a dust reddened continuum following from our earlier results. 

When comparing the emission lines in the three regimes, we again find that the two break strength $\lesssim$ 3 sub-populations share more similarities with each other than the break strength $>$ 3 regime. In the break strength $\lesssim$ 3 regimes, we observed a prominent (possibly broad) [\ion{C}{3}] emission that is comparatively weaker and blended with [\ion{He}{2}] in the break strength $>$ 3 regime. We find a strong, narrow Ly$\alpha$ emission feature in the break strength $>$ 3 regime, compared to a less prominent but clearly identifiable Ly$\alpha$ profile in the break strength $<$ 3 regimes. We identify strong, broad [\ion{Mg}{2}] emission in the break strength $>$ 3 regime that is not present in the other two regimes.

Our findings are consistent with the stacked composite spectra presented in Figure 8 of \citealt{Perez-gonzalez_26}, in which their sample is divided by the luminosity ratio between 5100 \AA\ and 2500 \AA. This parameter effectively traces the ``V-shaped'' continuum feature of LRDs; however, it does not directly correlate with the Balmer break strength (see the bottom panel of Figure 12 of \citealt{Perez-gonzalez_26}), where they demonstrate that the luminosity ratio remains flat with respect to the break strength, and only when the luminosity ratio exceeds 4 and the break strength exceeds 2 do the two parameters begin to correlate. While the classification scheme differs from ours, we observe the same qualitative trend: as the Balmer break becomes more prominent, the UV continuum weakens and the [\ion{Mg}{2}] becomes more prominent. Our composite spectra and classification reveal a more pronounced difference between UV continua of different break strength regimes, whereas \citet{Perez-gonzalez_26} find a more gradual reddening of the UV continuum across their classified populations. Their stacked spectra also show clearer evidence of [\ion{C}{3}] across the different sub-populations, which we find to be more prominent in the intermediate break strength and power-law SED LRDs, and blended in the extreme breaker (break strength $>$ 3) population.

\subsubsection{Rest Optical Properties of Composites}

The differences and similarities that we find at UV wavelengths change significantly when considering how the three regimes compare at rest-frame optical wavelengths. In the rest-frame optical, we instead find that the more pronounced difference occurs across the break strength $=$ 1 boundary, where the composites of LRDs with a Balmer break (break strength $>$ 1) show more similarities with each other than the break strength $<$ 1 composite. We find that the break strength $>$ 1 composites have a brighter optical continuum compared to their rest optical emission line strength. Since the composites are normalized around 0.8 ${\rm\mu m}$, another interpretation of the relative continuum brightness is that the power-law SEDs are intrinsically brighter in the red rest optical, consistent with previous results reporting a rising red rest optical in a subset of LRDs \citep{Delvechio_25, Barro_25}. We find that the broad rest-frame optical H$\gamma$, H$\beta$, [\ion{O}{3}] emission lines are strongest in the break strength $<$ 1 regime, with their relative strengths decreasing with increasing break strength. In particular, the [\ion{O}{3}] and H$\beta$ emission lines are significantly weaker in the break strength $>$ 3 composite relative to the other two composites. The broad H$\alpha$ emission line, however, appears comparable in all three composites.

A direct comparison of our rest-frame optical composite spectra with the stacked spectra of \citet{Perez-gonzalez_26} is complicated by the difference in normalization, as \citet{Perez-gonzalez_26} normalize their composites at 0.55$\mu$m. However, their composites with lower Balmer break strengths exhibit a rise in the red rest optical, consistent with the behavior we identify in the rest-frame optical here. Furthermore, our observation of the emission line properties are consistent with their findings, as both analyses show that $\Hb$ and [\ion{O}{3}] emission becomes weaker in LRDs with stronger Balmer breaks.

\subsubsection{Dust attenuation in the break strength $<$ 3 regime}

In this section, we explore the break strength $<$ 3 regime, where we have demonstrated that large scale dust is the dominant driver of both the break strength and the reddening of the UV emission. To investigate how this behavior manifests in the observed spectra, we define a new sub-population based on $\buv$ by dividing the population at a threshold of $\buv$ = -1.5, which we adopt as a boundary that effectively separates the two sub-populations into a dust attenuated ($\buv$ $>$ -1.5) and less-attenuated ($\buv$ $<$ -1.5) population. Composite spectra are then constructed for each sub-population as described in the previous section. 

Figure \ref{fig:composite_uv} presents the composite spectra of the two sub-populations, normalized over the wavelength range 0.77--0.83\ $\mu$m, following the approach of Figure \ref{fig:composite_all_three}. The attenuated population ($\buv$ $>$ -1.5) exhibits a systematically redder UV continuum relative to the unattenuated population, while the rest optical continuum of the two populations are remarkably similar. This behavior is fully consistent with the trends identified in Section \ref{sec:narrow_line}, and provides a striking spectroscopic illustration of the decoupling between the UV and rest-frame optical continua in dust-attenuated LRDs, highlighting the divergence of the dust behavior in these LRDs from a simple screen geometry.

\section{Discussion}
\label{sec:conclusions}
\subsection{Origin of the UV Emission in LRDs}

In our analysis, we measure the narrow-line Balmer decrement of LRDs and identify a correlation (Figure \ref{fig:bdec}) between the narrow-line Balmer decrement and the UV continuum slope, $\buv$, which we then interpret as evidence that large-scale dust attenuation drives the reddening of the UV emission in LRDs. This interpretation is further reinforced by SED fitting analysis that reveals that SFGs and LRDs may share similar intrinsic stellar templates, with the observed redness of the LRD $\buv$ compared to the SFGs being primarily  driven by stellar mass and stellar attenuation. However, it remains unclear if this interpretation holds for all LRDs, as the SED fitting of the LRD UV continuum shows that for LRDs with the reddest $\buv$ the stellar template is not able to adequately fit the UV emission alone, raising the question of whether there is another emission component in the UV in LRDs.

The origin of the UV emission in LRDs remains an open question. Some studies suggest it arises from the host galaxy \citep{Rinaldi_2025, Naidu_25, Volonteri2025}, while others argue it is dominated by scattered emission from a broad line AGN enshrouded by dense gas, a black hole star \citep{Greene_26, Pacucci_26}. \citet{Volonteri2025} introduces a framework in which a host-galaxy-dominated UV continuum and an AGN-dominated rest-frame optical continuum can reproduce the photometric color criteria used to select LRDs. Within this framework, they predict a median stellar dust attenuation of $A_{\rm V}$ $\sim$ 0.8 arising from dust clumps around star-forming regions, and a nuclear attenuation of $A_{\rm V}$ $\sim$ 2.4 required for the host galaxy to dominate the rest-frame UV emission. Alternatively, \citet{Ji_2025} fit the UV continuum of an LRD with both a gas attenuated AGN component and a nebular continuum, finding that the required dust attenuation for the AGN component is $A_{\rm V} \sim 0$ in both models considered. 

Our interpretation is consistent with the host galaxy scenario proposed by \citet{Volonteri2025}, and our inferred range of attenuation broadly agrees with their predicted median $A_{\rm V}$. While it remains unclear whether this applies uniformly across the full LRD population, and whether LRDs constitute a homogeneous population of galaxies, we find that for the subset of LRDs with moderate Balmer break strengths, large-scale dust attenuation is the most likely driver of the observed UV reddening.

The presence of broad UV emission lines such as \ion{C}{3}] and \ion{Mg}{2}] in the composite spectrum in Figure \ref{fig:composite_all_three} suggests the AGN may also contribute to the UV emission, as these lines are typically used as AGN diagnostics in local AGN \citep{Netzer_90}. Recently, \citet{Ando2026} from a similar analysis to ours, comparing the UV emission between the LRDs and SFGs argued that the UV emission in LRDs can not be explained by simply invoking a host galaxy, and they further argued that the UV emission requires contribution from a central compact and red source. While we do find a similar result here, we show that this conclusion does not hold for the full populations of LRDs but rather for a subset of sources exhibiting both a red $\buv$ and strong ( $\geq$ 2) Balmer break. Additionally, \citet{Ji2026} reported the detection of broad Ly$\alpha$ emission in the far-UV spectrum of an LRD at z$=$7.04, interpreted as evidence for a non-unity covering fraction of the gas enshrouding the central BLR, through which scattered AGN emission is able to escape. However, recent MUSE spectroscopy analysis of the Ly$\alpha$ emission profile in LRDs constrains the broad Ly$\alpha$ fraction to $\sim$ 10\%, suggesting that the contribution from a central compact source to the UV emission in LRDs may not be as common as previously assumed \citep{Kageura2026}. While the presence of these broad UV emission lines is suggestive of AGN activity, it remains unclear whether classical AGN line diagnostics developed for local, solar-metallicity populations can be reliably applied to high-redshift, low-metallicity systems \citep{Ubler_2023}, and previous attempts to characterize LRDs based on \ion{C}{3}] emission have proven inconclusive \citep{Perez-gonzalez_26}. However, the combination of the broad UV emission lines and the redder $\buv$ values in LRDs compared to SFGs suggests that scattered AGN emission may contribute to the overall UV emission in LRDs. 

Lastly, one aspect we do not probe in our analysis, which could provide further constraints on whether an additional emission component is required to explain the UV continuum in LRDs, is whether alternative IMFs could result in a better fit to the UV continuum, particularly for the sub-population with redder $\buv$ values. We reserve this analysis for a follow-up paper, as it lies beyond the scope of our question here, which was simply whether a dust attenuated stellar template alone can fit the UV continuum in LRDs.

\subsection{Origin of the Balmer break in LRDs}

There is a growing consensus in the literature regarding the origin of the Balmer break feature in LRDs with extreme Balmer breaks. These ``extreme breaker'' LRDs exhibit Balmer break strengths that exceed what is theoretically achievable from stellar bound-free absorption alone, leading to the attribution of this feature to absorption by the dense gas envelope in the enshrouded black hole model \citep{Naidu_25, DeGraff_2025, juodzbalis_24}. However, for LRDs with more moderate Balmer breaks, it remains unclear whether the feature arises primarily from stellar absorption, from absorption by the gas envelope, or from a combination of both.

We identify a clear correlation between the narrow-line Balmer decrement and the Balmer break strength in LRDs (Figure \ref{fig:bdec}). The break strength values of the LRDs analyzed here are broadly consistent with the theoretical range achievable from dusty, evolved stellar populations, similar to those observed in dusty starburst galaxies \citep{Kriek_2006}. Furthermore, our SED fitting analysis shows that for LRDs with moderate to weak Balmer break strengths ($<2$), the break feature can be fit with a single stellar template alone (See Figure \ref{}). However, for the LRDs with stronger Balmer break values, the stellar template is not able to fit the UV continuum while also reproducing the observed Balmer break. An additional effect observed from the SED fitting of the LRDs with the strong Balmer breaks is that the best-fit stellar mass is always too high ($\log(\Mstar/M_{\odot})$ $\geq$ 10), which have been shown to be highly unlikely given the typical effective radius of LRDs ($r_e$ $\sim$ $10^{1-2}$ pc), as the stellar densities implied would exceed maximum theoretical and observed densities of star clusters and galactic nuclei \citep{de_graff_25a}. This further reinforces the idea that the Balmer break in these strong breakers can not simply be attributed to a stellar origin, as the SED fitting pushes the stellar template to the most extreme regimes while still failing to produce a good fit to the feature.

We also identified Balmer absorption features in three of the 13 LRDs in our sample. The Balmer absorption feature observed in LRDs likely arises from collisionally excited $n=2$ neutral hydrogen in dense neutral gas clumps, and has been suggested to be physically linked with the Balmer break, as both may originate from the same dense neutral gas \citep{Inayoshi_Maiolino_2025}. However, it remains unclear whether these two features are indeed physically connected; \citet{Matthee2026} report a correlation between the Balmer break strength and the velocity offset of the absorber as evidence of a common physical origin, while \citet{Lin_2026} find no such correlation for the same two parameters.

Regardless of whether the origin of the break strength is stellar or nuclear, or some combination of both, our analysis demonstrates that the emitting region responsible for the Balmer break likely resides behind the same large-scale dust screen that reddens the UV emission, as evidenced by the tight correlation between the Balmer break strength and the narrow-line Balmer decrement. 

\subsection{Implications for LRD Geometry}

The unique combination of an attenuated UV continuum, an attenuated Balmer break, and an unattenuated rest-frame optical continuum as shown in Figure \ref{fig:bdec} suggests that the multi-wavelength emission in LRDs has multiple distinct physical origins. The UV emission likely originates primarily in the host galaxy, while the rest-frame optical continuum arises from the central AGN in LRDs. The dust attenuation behavior observed in our analysis further suggests a specific geometric configuration for LRDs. If the narrow-line Balmer decrement traces large-scale dust attenuation, as we have argued here, and if the nuclear emission, comprising the broad emission lines and the rest-frame optical continuum, were seen through the same dust screen, one would expect the attenuation signature to be observed uniformly across all wavelengths. However, this is not what we observe. As shown in Figure \ref{fig:composite_uv}, comparing the two sub-populations selected based on $\buv$, the UV continuum is consistent with dust attenuation for the redder UV population, while the rest optical continua for the two sub-populations lie essentially on top of one another. Similarly, we show in Figure \ref{fig:bdec} that while $\buv$ is broadly correlated with the narrow-line Balmer decrement, $\bopt$ appears to be scattered with no correlation. We therefore argue that the observed behavior in the SED of LRDs with respect to large scale dust attenuation is consistent with a face-on geometric configuration in which the line of sight to the nucleus (dominating the optical) passes through relatively little dust, while the star-forming regions in the host galaxy (dominating the UV) remain significantly dust-attenuated.

\section{Summary}
\label{sec:summary}

In this paper, we present an analysis of 89 NIRSpec PRISM LRDs, with a focused emission line fitting analysis of a subset of 13 LRDs with medium-resolution (G235M and G395M) grating spectroscopy. We perform a two-component broad and narrow line decomposition to investigate the presence and impact of large-scale dust attenuation on the spectral shapes of LRDs. Our main results are summarized as follows:
\begin{itemize}
    \item The narrow-line Balmer decrements of LRDs are elevated relative to the case B recombination value of 2.86, and systematically higher than those measured for 5$<$z$<$6 star-forming galaxies. The majority of LRDs in our sample have narrow-line Balmer decrement values raging from 3 to 13, corresponding to an $E(B-V)$ range of 0.04 - 1.3 \citet{Calzetti_2000} $E(B-V)$, revealing moderate to significant large-scale dust attenuation presence in the LRDs.
    
    \item The narrow-line Balmer decrement is strongly correlated with both $\buv$ and the Balmer break strength, indicating that large-scale dust attenuation plays a significant role in governing the UV continuum slope and the prominence of the Balmer break in LRDs. In contrast, the rest-frame optical continuum slope $\bopt$ shows no systematic trend with the narrow-line Balmer decrement, suggesting that the rest-frame optical continuum is not significantly affected by the same large-scale dust attenuation.
    
    \item Classifying LRDs by Balmer break strength reveals a sub-population of LRDs with break strength $\lesssim$ 3 in which large-scale dust attenuation is likely the dominant driver of both the UV continuum slope and the optical spectral shape. SED fitting analysis shows that origin of the Balmer break in LRDs with moderate break strength ( $<$ 2) is consistent with a dust attenuated stellar template, while the stronger break LRDs are inconclusive. Additionally, for the strong break LRDs, it remains unclear whether large scale dust attenuation is present and what role it plays.
    
    \item Stacked composite spectra of LRDs classified by Balmer break strength reveals broad emission that could be evidence of AGN contribution to the rest-frame UV emission. Additionally, composites reveal systematic decrease in the strength of rest-frame optical emission lines with increasing break strength, with an especially prominent decrease in the \ion{O}{3} emission strength for the strong break LRDs.
\end{itemize}

\vspace{10pt}
\texttt{Acknowledgments:} 
E.D. acknowledges support from the Dartmouth graduate fellowship. R.C.H. acknowledges support from NASA ADAP grant number 80NSSC25K7569. This research made use of Astropy, a community-developed core Python package for Astronomy \citep{2018AJ....156..123A, 2013AA...558A..33A},  \texttt{OverCite} \citep{Shariat2026}, an in-editor citation tool for \LaTeX, pandas \citep{McKinney_2010, McKinney_2011}, and matplotlib, a Python library for publication quality graphics \citep{Hunter:2007}. 

\texttt{Data Availability:}
The data used in this work are publicly available through the
DAWN JWST Archive (https://dawn-cph.github.io/dja/index.html).

\appendix
\counterwithin{figure}{section}

\section{Line Fitting Results and MCMC PDF \label{sec:appendix}}
We present the full results of the Gaussian fitting analysis for all 13 LRDs with medium-resolution spectroscopy.

\begin{figure*}[!h]
    \centering
    \includegraphics[width=1\linewidth]{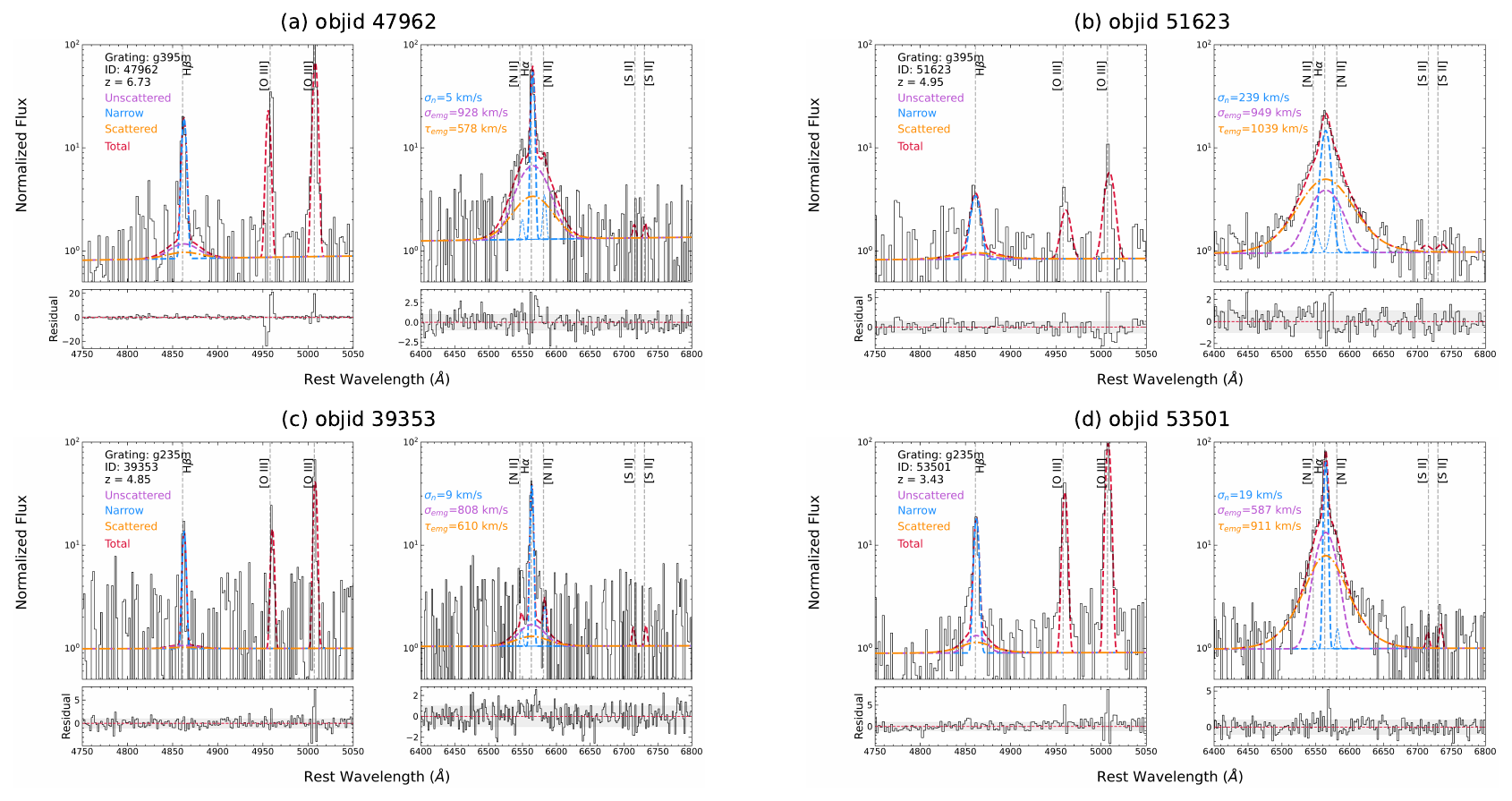}
    \caption{Two-component Gaussian + exponential fitting results similar to Figure \ref{fig:cutout} for LRDs:  EGS-NELSONX-47962, EGS-NELSONX-51623, JADES-GDN-39353, and ADES-GDN-53501. $\Hb$ is on the left panel while $\Ha$ is shown on the right panel. The orange dashed line depicts the scattered emission (an intermediate Gaussian convolved with a symmetric exponential), the purple dashed line depicts the unscattered intermediate Gaussian, the blue dashed line depicts the narrow emission, the black dashed line depicts the narrow Balmer absorption, and the red dashed line depicts the total model fit to the data. Vertical dashed line markers indicate the positions of the Balmer emission lines alongside the forbidden emission lines included in the fitting analysis: [\ion{O}{3}] doublet, [\ion{N}{2}] doublet, and [\ion{S}{2}] doublet. The residuals have also been added in a separate panel. The best-fit values for the different velocity width of each line comp are indicated. }
    \label{fig:full_fit_1}
\end{figure*}
\begin{figure*}[!h]
    \centering
    \includegraphics[width=1\linewidth]{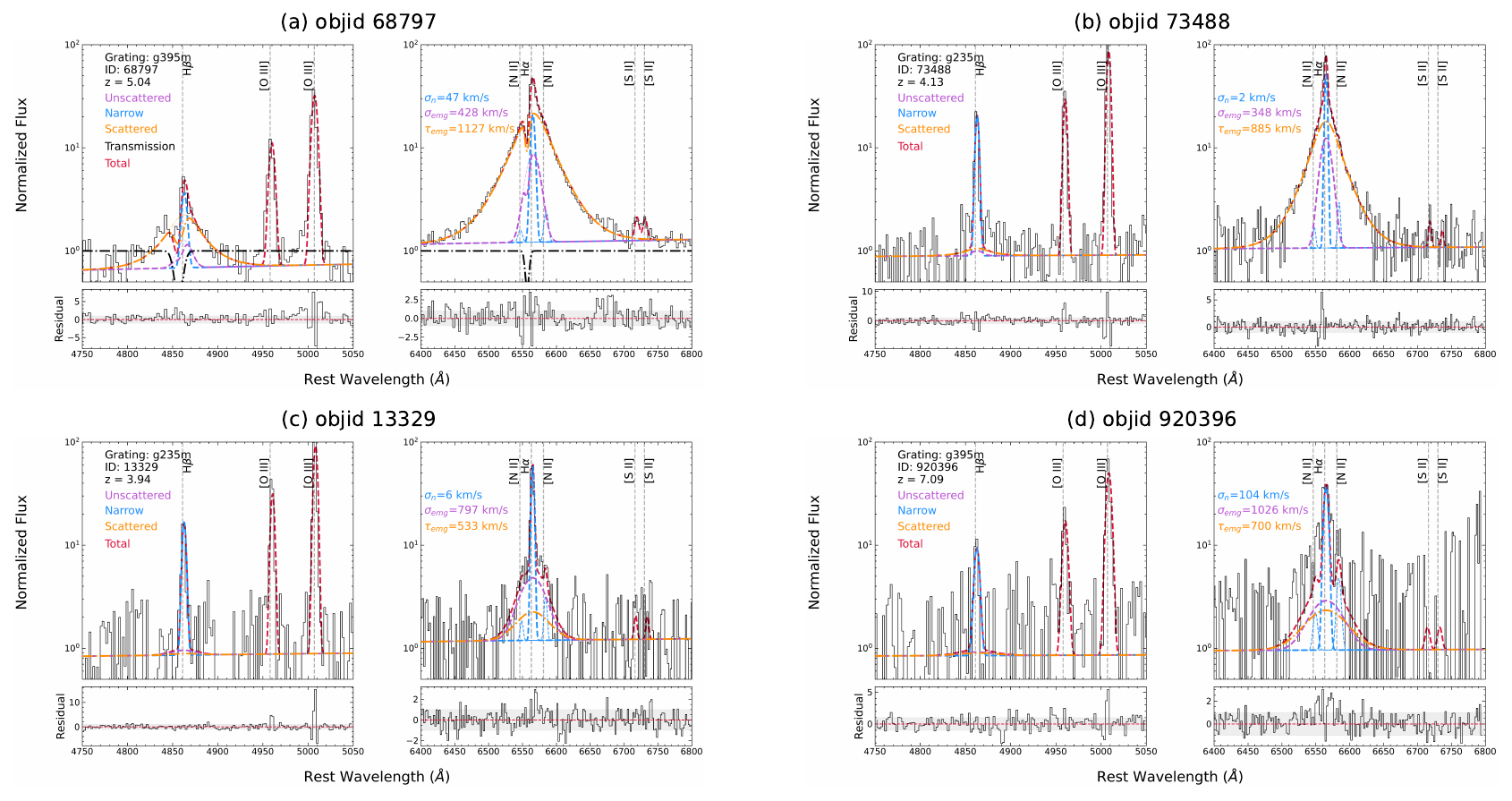}
    \caption{Same as in \ref{fig:full_fit_1} for JADES-GDN-68797, JADES-GDN-73488, JADES-GDS-13329, and MOM-UDS-920396. }
    \label{fig:full_fit_2}
\end{figure*}
\begin{figure*}[!h]
    \centering
    \includegraphics[width=1\linewidth]{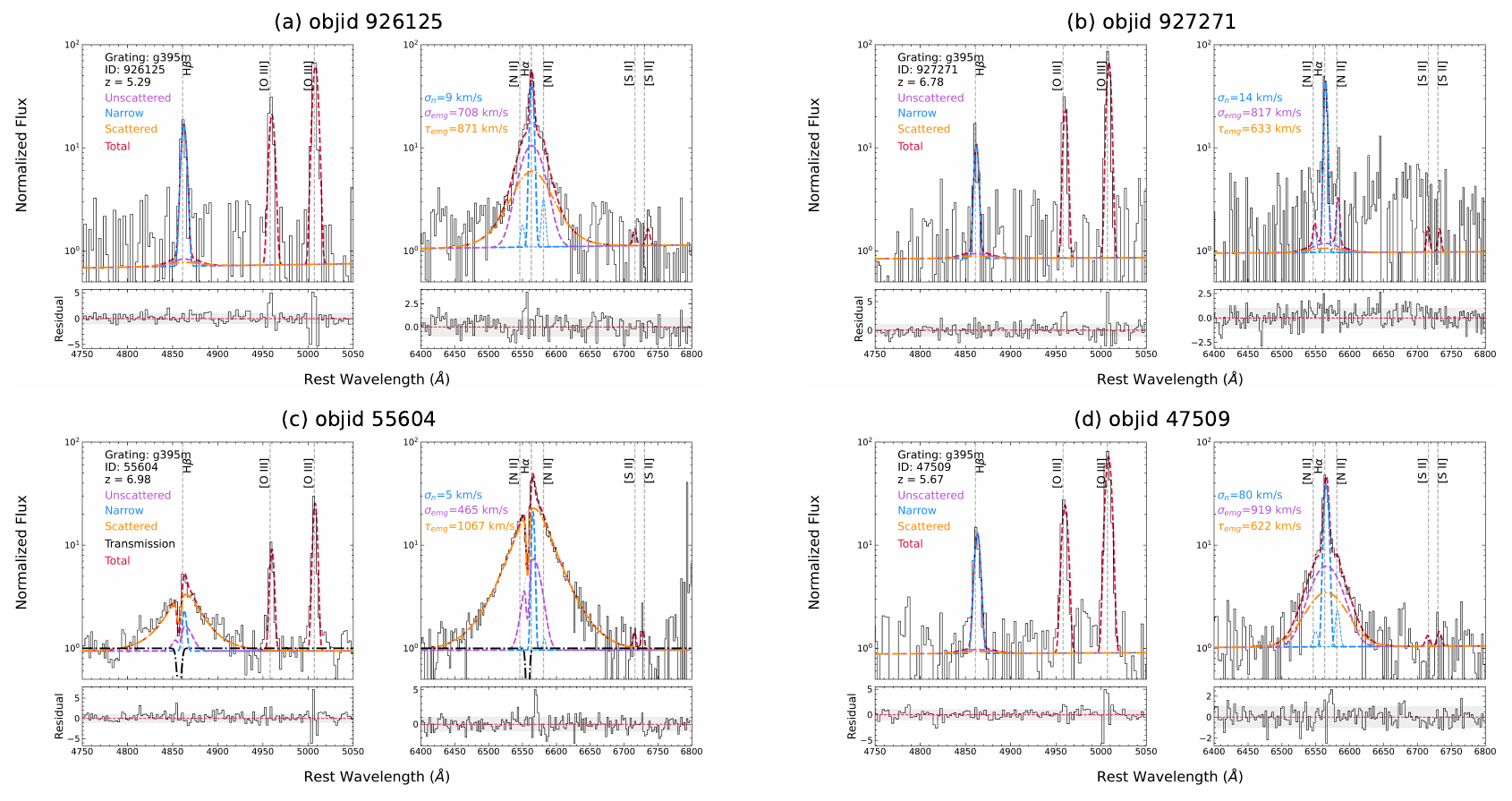}
    \caption{Same as in \ref{fig:full_fit_3} for RUBIES-EGS-926125, RUBIES-EGS-927271, RUBIES-EGS-55604, and RUBIES-EGS-47509.  }
    \label{fig:full_fit_3}
\end{figure*}

\bibliography{ms}
\bibliographystyle{aasjournalv7.1}



\end{document}